\documentclass[twocolumn,showpacs,preprintnumbers,amsmath,amssymb]{revtex4}
\usepackage{graphicx}
\usepackage{dcolumn}
\usepackage{bm}
\usepackage{longtable}
\usepackage{tabularx}
\begin{document}

\preprint{}

\title{
Variation after parity projection calculation
with Skyrme interaction for light nuclei
}

\author{H. Ohta}
\author{K. Yabana}
\affiliation{Institute of Physics, University of Tsukuba, 
Tsukuba 305-8571, Japan}

\author{T. Nakatsukasa\footnote{Present address:
Institute of Physics, University of Tsukuba,
Tsukuba 305-8571, Japan}}
\affiliation{Department of Physics, Tohoku University,
Sendai 980-8578, Japan}

\date{\today}

\begin{abstract}
A self-consistent calculation with variation 
after parity projection is proposed to study both ground
and excited states of light nuclei. 
This procedure provides description of the ground state incorporating
some correlation effects, and self-consistent solutions for
the excited states of negative parity.
For flexible description of nuclear shapes, single particle 
orbitals are represented on a uniform grid in the three-dimensional 
Cartesian coordinates. The angular momentum projection is 
performed after variation to calculate rotational spectra.
To demonstrate the usefulness of the method, results are shown
for two $N=Z$ nuclei, $^{20}$Ne and $^{12}$C, for which
clustering correlations are known to be important.
In the $^{20}$Ne nucleus, both cluster-like and shell-model-like
states are described simultaneously in the present framework.
For $^{12}$C nucleus, the appearance of three-alpha clustering 
correlation in the ground state is investigated in relation to 
the strength of the two-body spin-orbit interaction.
\end{abstract}

\pacs{21.10.Hw, 21.60.Jz, 27.20.+n, 27.30.+t}
\maketitle

Nuclear mean-field calculations have been successful for 
systematic description of nuclear ground state properties 
with a few adjustable parameters \cite{Bender03}. The mean-field theory
has also been extensively applied to description of excited states.
For example, collective vibrational as well as single 
particle-hole excitations are described by the random phase 
approximation which is equivalent to a small amplitude limit 
of the time-dependent mean-field dynamics. Excited states may also
be described as local-minimum solutions if there is more than 
one self-consistent solution in the static calculation.
In principle, the generator coordinate method is expected 
to provide a unified description for any kind of excitations. 
However, in practice, one need to specify a few important
generator coordinates in advance by physical intuition.

In this article, we present an attempt to describe both ground
and excited states in the self-consistent approach 
with the variation after parity projection (VAPP).
For an even-even nucleus, the lowest energy
positive-parity solution describes the ground state in which 
the correlation beyond the mean field is incorporated to a certain extent. 
Self-consistent negative-parity solutions should correspond to
the negative-parity excited states. 

The theory of the variation after projection (VAP) of
the symmetry-violating internal state
has a long history \cite{Ring80}. However, practical applications 
with full variation of single-particle orbitals are rather few
even for projection with respect to the parity.
Previously we achieved the VAPP
calculation employing the uniform grid representation in the 
three-dimensional (3D) Cartesian coordinates in which the simple 
BKN interaction was used \cite{Takami96}.
In the ordinary mean-field calculations, one usually obtains
self-consistent solutions with axial and reflection symmetries
for most nuclei. However, in the VAPP
calculations, the self-consistent solutions are found to
violate these symmetries. A preliminary report on the realistic 
calculations employing 
the Skyrme interaction has been presented in Refs.~\cite{Takami97,Takami98}. 

We apply the method to light $N=Z$ nuclei, $^{20}$Ne and $^{12}$C.
There are several reasons why we consider the light nuclei.
For light nuclei, the VAP may significantly modify the mean-field
solutions and lead to the energy gain by the projection.
In the excited states of light nuclei, various cluster structures
have been found to appear. They have been successfully analyzed
with the microscopic cluster models in which the VAP calculation
was carried out \cite{Fujiwara80}. The present approach 
is useful to understand the mechanism for the 
appearance of the cluster structures. Finally, there are rapid 
progresses in the experimental research on light unstable nuclei.
The present framework will be useful to analyze and predict
structures of unstable nuclei.

The antisymmetrized molecular dynamics (AMD) method developed 
by Kanada-En'yo and Horiuchi \cite{Kanada01} has been utilized
to describe clustering phenomena in light stable and unstable nuclei. 
In the simplest version of AMD, the method is identical to the VAPP
with a restricted Slater determinant in which each orbital 
is a Gaussian wave packet. This may be regarded as
an approximation to our present approach. The feasibility of the
AMD allows them to perform the 
variation after angular momentum projection \cite{Kanada98}.
On the other hand, the present work brings more accurate description 
of the single particle orbitals, and establishes an intimate link to 
the Skyrme Hartree-Fock theory.

We start with a brief description of the theory \cite{Takami96}. 
We consider an arbitrary Slater determinant 
$\Phi=\frac{1}{\sqrt N!}{\rm det}\{\phi_i(x_j)\}$,
with $x=({\vec r},\sigma)$, the space and spin coordinates. 
We apply the parity projection operator for this Slater determinant,
$\Phi^{(\pm)}=\frac{1}{\sqrt{2}}(1 \pm {\hat P})\Phi$, 
where $\hat P$ is the space inversion operator. 
We then consider the energy expectation value for the states
with definite parity, $\Phi^{(\pm)}$, and make a variation with 
respect to the single particle orbitals $\phi_i$,
\begin{equation}
\frac{\delta}{\delta \phi_i^*} \left[ \frac{\langle\Phi^{(\pm)} \vert
{\hat H} \vert
\Phi^{(\pm)} \rangle} {\langle \Phi^{(\pm)} \vert \Phi^{(\pm)} \rangle}
- \sum_{i,j} \epsilon_{ij} \langle \phi_i \vert \phi_j \rangle 
- {\vec \eta} {\sum\limits_i} \langle \phi_i 
\vert {\vec r} \vert \phi_i \rangle \right]=0
\label{E-func}
\end{equation}
Here we imposed two kinds of constraints in the above variation
\cite{Takami96}.
The first one with the Lagrange multipliers $\epsilon_{ij}$ is 
introduced to orthonormalize single particle orbitals, $\phi_i$.
The second with the multiplier $\vec \eta$ is
to coincide the center-of-mass of the wave function $\Phi$ 
with the origin of the coordinate, 
$\sum_i \langle \phi_i \vert \vec r \vert \phi_i \rangle = 0$.
This constraint assures that the parity operation is made with respect
to the center-of-mass of the nucleus and minimizes the
occurrence of the spurious center-of-mass excitations by the parity 
projection.

The variation of Eq.~(\ref{E-func}) yields the following 
self-consistent equation,
\begin{eqnarray}
&&\left( h- \vec{\eta} \cdot \vec{r} \right) \phi_i \pm
\langle \Phi \vert {\hat P} \vert
\Phi \rangle \{h_P \tilde{\phi}_i - \sum_j \tilde{\phi}_j
\langle \phi_j \vert h_P \vert \tilde{\phi}_i \rangle \} \nonumber\\
&&~~~~~~+( E^{(\pm)} - E ) \tilde{\phi}_i = \sum_j \epsilon_{ij} \phi_j
\label{PPSHFeq}
\end{eqnarray}
where $h$ is the usual Hartree-Fock Hamiltonian. $h_P$ has
the same structure as $h$, however, all the densities are replaced
with the transition densities which are the matrix elements of
density operators between the wave function $\Phi$ and its 
parity-inverted state $\hat P \Phi$. $E^{(\pm)}$ is the energy
expectation value with respect to the wave function $\Phi^{(\pm)}$, 
while $E$ is with respect to $\Phi$. $\tilde \phi_i$ is defined by
\begin{equation}
\tilde \phi_i = \sum_j \hat P \phi_j (B^{-1})_{ji}
\end{equation}
with $B_{ij}=\langle \phi_i \vert \hat P \vert \phi_j \rangle$.

In the practical calculations, Eq.~(\ref{PPSHFeq}) is not solved
directly. Instead, the imaginary-time method is employed in which
the left-hand-side of Eq.~(\ref{PPSHFeq}) is used as the gradient
of the energy functional. We discretize the 3D Cartesian coordinates
into uniform square grid and represent the single particle wave functions
on the grid points. The grid spacing is taken to be 0.8 fm,
and the grid points inside a sphere of radius 7.2 fm are used in the 
calculations below.

After obtaining the self-consistent solutions, we make the angular
momentum projection (AMP) to calculate rotational spectra.
The self-consistent solutions in the VAPP
are usually not axially symmetric. Therefore, 
one must perform three-dimensional rotation in
Euler angles, $\Omega=(\alpha,\beta,\gamma)$, for the AMP. 
We define the AMP state as usual,
\begin{equation}
\vert \Phi^{J(\pm)}_{MK} \rangle = \frac{2J+1}{8\pi^2} \int d\Omega 
D^{J~*}_{MK}(\Omega) R(\Omega) \vert \Phi^{(\pm)} \rangle
\label{JPwf},
\end{equation}
where $R(\Omega)$ is the rotation operator and $D^J_{MK}(\Omega)$
is the Wigner's $D$-function.
We then define relevant matrix elements, $h^{J(\pm)}_{KK'}$ and
$n^{J(\pm)}_{KK'}$. The Hamiltonian matrix element $h^{J(\pm)}_{KK'}$ is
given by
\begin{eqnarray}
\label{Angle_Integral}
h^{J(\pm)}_{KK'} &=& 
\langle \Phi^{J(\pm)}_{MK} \vert 
\hat H \vert \Phi^{J(\pm)}_{MK'} \rangle
 \nonumber\\
 &=& \frac{2J+1}{8\pi^2} \int d\Omega D^{J~*}_{KK'}(\Omega)
\nonumber\\ 
&& \times
 \langle \Phi \vert e^{-i\alpha \hat J_z}
 \hat H (1\pm \hat P) e^{-i \beta \hat J_y}e^{-i \gamma \hat J_z}
 \vert \Phi \rangle,
\end{eqnarray}
and the similar expression for the norm matrix element, $n^{J(\pm)}_{KK'}$.
For practice, we achieved the rotation of the 
single particle orbitals by successive rotations of a small angle.
Following the Taylor expansion method for time evolution which was
employed in the time-dependent Hartree-Fock calculations \cite{Flocard78},
we calculate the rotation of small angle $\Delta \alpha$ 
around $z$-axis by
\begin{equation}
\phi_i^{\alpha+\Delta \alpha} 
= \sum_{k=0}^{N_{max}}
\frac{(\Delta \alpha)^k}{k!} (j_z)^k \phi_i^{\alpha},
\end{equation}
where $N_{max}$ is taken to be 4.
Typically, each Euler angle is discretized into 200 steps for
$\alpha$ and $\gamma$, and into 400 steps for $\beta$.
In Eq.~(\ref{Angle_Integral}), the integrations over Euler angles
are achieved with discretization of 20 points for $\alpha$ and $\gamma$ and
400 (all) points for $\beta$. It is important to divide rotations
of three Euler angles into two and one; the rotations of the angles 
$\beta$ and $\gamma$ for the ket state, and the
rotation of the angle $\alpha$ for the bra state, 
to reduce the computational costs.

We employ the Skyrme interaction in constructing the energy
functional in Eq.~(\ref{E-func}).
There is ambiguity in the choice of the density of density dependent
force for off-diagonal matrix elements \cite{Duguet03}, which appear 
in Eqs.~(\ref{E-func}), (\ref{PPSHFeq}), and (\ref{Angle_Integral}).
In this paper, we simply use corresponding transition densities for them.
We will hereafter abbreviate the present scheme as the parity-projected 
Skyrme Hartree-Fock (PPSHF) method.
 
Before showing our results, we mention some
problematic aspects of the Skyrme interaction in the present 
framework which we have encountered in the practical 
calculations. Since the Skyrme interaction has been so constructed 
to be used for a single Slater determinant state,
it is by no means evident whether it may provide a reasonable 
description in the VAPP calculations.
Indeed, we have observed that, if we make a variation of single 
particle orbitals without any restriction, we obtain an unphysical 
solution in which the time-reversal symmetry is violated in the 
ground state. For example, the total binding energy of $^{20}$Ne 
in the ordinary Skyrme Hartree-Fock (SHF) calculation with SGII 
force is about 170 MeV, while we obtain the binding energy 
of 230 MeV with the PPSHF. Since the parameterization of the Skyrme 
interaction relevant to the time-odd component has not been fully
tested, we restrict our wave function to be time-reversal invariant.
Namely, we force nucleons either to fully occupy or unoccupy 
a pair of orbitals which are mutually 
related to each other by the time-reversal operation in the Slater
determinant. This restriction removes most part of 
the problems. However, in the positive parity solutions of some 
light nuclei, we still encounter a problem that the density of 
the self-consistent solution shows a small but unphysical 
oscillation, namely, the density shows staggering in the 
neighboring grids. This problem can be overcome if we further 
ignore a part of the Skyrme energy functionals including terms
$\vec \rho$, $\vec j$, $\vec T$ \cite{Engel75, Bonche87}. 
For a single Slater determinant with 
time-reversal invariance, these time-odd densities, 
$\vec \rho$, $\vec j$, $\vec T$,  vanishes identically.
However, this is not the case for transition densities between 
$\Phi$ and $\hat P \Phi$.

The problems are related to the fact that
the superposition of two Slater determinants by the parity
projection may represent more varieties of correlation effects 
than the single Slater determinant may do. We leave it an 
interesting future problem to find appropriate Skyrme energy
functional to be used in the PPSHF framework.
In the following calculations, we will employ SGII
parameter set \cite{Giai81}. 
This force has been successful in describing the 
response properties of nuclei, although the absolute binding
energies are not reproduced accurately.

\begin{table}[tb]
\caption{
\label{tab:20Ne}
Self-consistent solutions of $^{20}$Ne are summarized. 
The third row (``Energy (MF)'') shows the results of ordinary SHF and 
PPSHF ($K^{\pi}=0^+$, $K^{\pi}=2^-$, $K^{\pi}=0^-$) before
AMP. Total energy is shown for the ground state (SHF and
$K^{\pi}=0^+$) , while the excitation energies with respect 
to the ground state are shown for the excited states
($K^{\pi}=2^-$ and $K^{\pi}=0^-$).
The fourth raw (``Energy (MF+AMP)'')
indicates the results for the lowest angular momentum
state after the AMP. The fifth raw (``$E_{LS}$'')
is the expectation value of the two-body spin-orbit interaction. 
The $\beta_2$, $\gamma$, $\beta_{30}$ and $\beta_{32}$, 
are the density deformation
of the internal single Slater determinant.
}
\begin{tabular}{l|cccc}
\hline\hline
 & $K^\pi=0^+$ & $K^\pi=2^-$ & $K^\pi=0^-$
 & SHF \\
\hline
Energy (EXP)    & $-160.645$ & $+4.97$    & $+5.78$   & $-160.645$ \\
\hline
Energy (MF)     & $-168.232$ & $+5.47$    & $+7.96$   & $-167.943$ \\
Energy (MF+AMP) & $-171.333$ & $+4.91$    & $+6.42$   &            \\
${\rm E_{LS}}$  & $-8.517$   & $-15.350$  & $-5.443$  & $-8.775$   \\
$\beta_2$       & $0.535$    & $0.589$    & $0.694$   & $0.728$    \\
$\gamma$        & $0.0$      & $0.0$      & $0.0$     & $0.0$      \\
$\beta_{30}$    & $0.314$    & $0.0$      & $0.576$   & $0.0$      \\
$\beta_{32}$    & $0.0$      & $0.178$    & $0.0$     & $0.0$      \\
\hline\hline
\end{tabular}
\end{table}

We first show the results of $^{20}$Ne.
In Table~\ref{tab:20Ne}, the calculated results are summarized.
The definition of the deformation parameters is in Ref.~\cite{Takami98-2}.
Octupole deformation parameters $\beta_{31}$ and $\beta_{33}$ are
negligible and not shown.
The positive-parity solution corresponds to the ground state. 
In the SHF calculation, this has a prolate shape
with reflection symmetries. On the other hand, the PPSHF 
calculation gives the solution with substantial $\beta_{30}$ 
deformation. The energy in the PPSHF calculation gets lower by 
0.3 MeV from the ordinary SHF calculation.

There are two solutions in the negative parity. The lowest energy 
solution has a small $\beta_{32}$ deformation as well as the prolate 
deformation. The next solution has a large $\beta_{30}$ deformation
as well as $\beta_2$. In the angular momentum projection, we 
found that the former solution is characterized by $K^{\pi}=2^-$ and 
the latter solution by $K^{\pi}=0^-$.

Previously we reported the existence of two local minima in the
negative parity \cite{Takami97}. After careful examination, we have 
found that the $K^{\pi}=0^-$ solution is not a local minimum solution
but decays into the $K^{\pi}=2^-$ solution after very long 
imaginary-time iterations (typically $5 \times 10^3$ steps with 
$\Delta t = 10^{-3}$ MeV$^{-1}$).
Reflecting different $K^{\pi}$ value of these two states, 
they are almost orthogonal to each other. To examine the physical
reality of the $K^{\pi}=0^-$ solution, we have performed the 
imaginary-time evolution of this solution with an extra 
constraint that the wave function is orthogonal to the 
$K^{\pi}=2^-$ solution. 
In this procedure, we obtain a converged solution which is found
to be almost the same as the original $K^{\pi}=0^-$ quasi-stable 
solution without the orthogonalization.
Because of these observations and a good correspondence to the 
measured spectra as shown below, we consider that the $K^{\pi}=0^-$ 
solution is of physical significance.

\begin{figure}[tb]
\begin{minipage}{.10\linewidth}
(a)
\end{minipage}
\begin{minipage}{.24\linewidth}
\includegraphics[width=\linewidth]{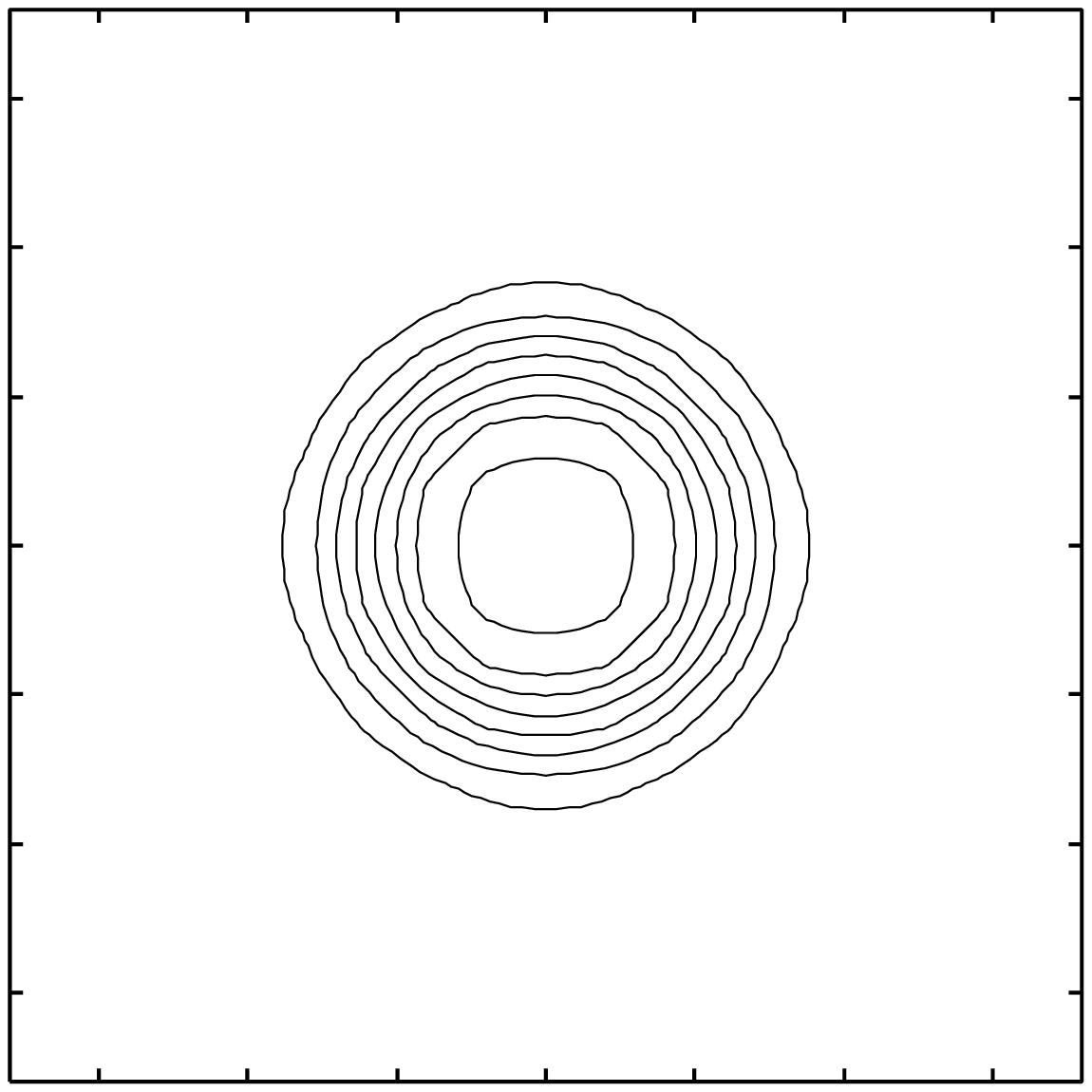}
\end{minipage}
\begin{minipage}{.24\linewidth}
\includegraphics[width=\linewidth]{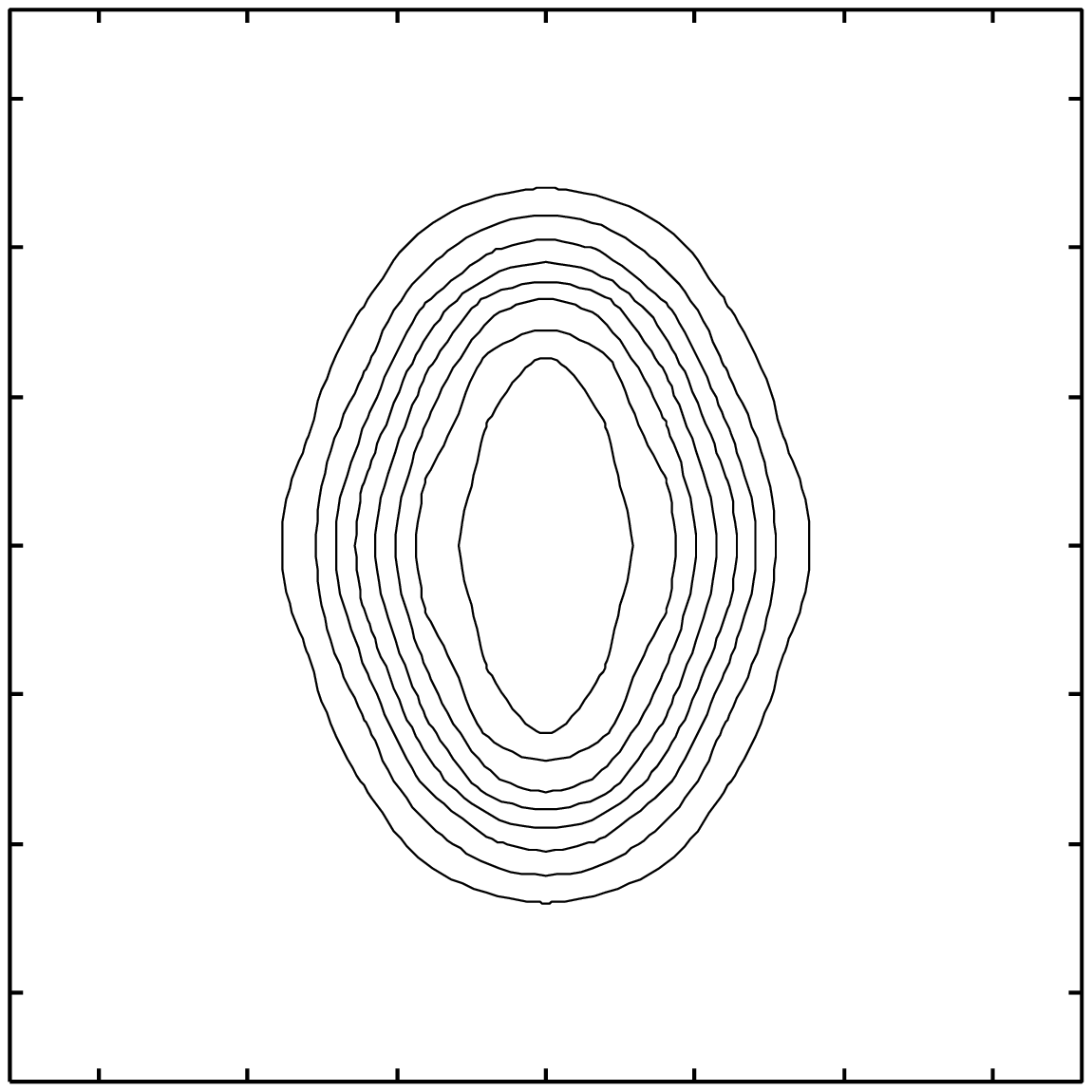}
\end{minipage}
\begin{minipage}{.24\linewidth}
\includegraphics[width=\linewidth]{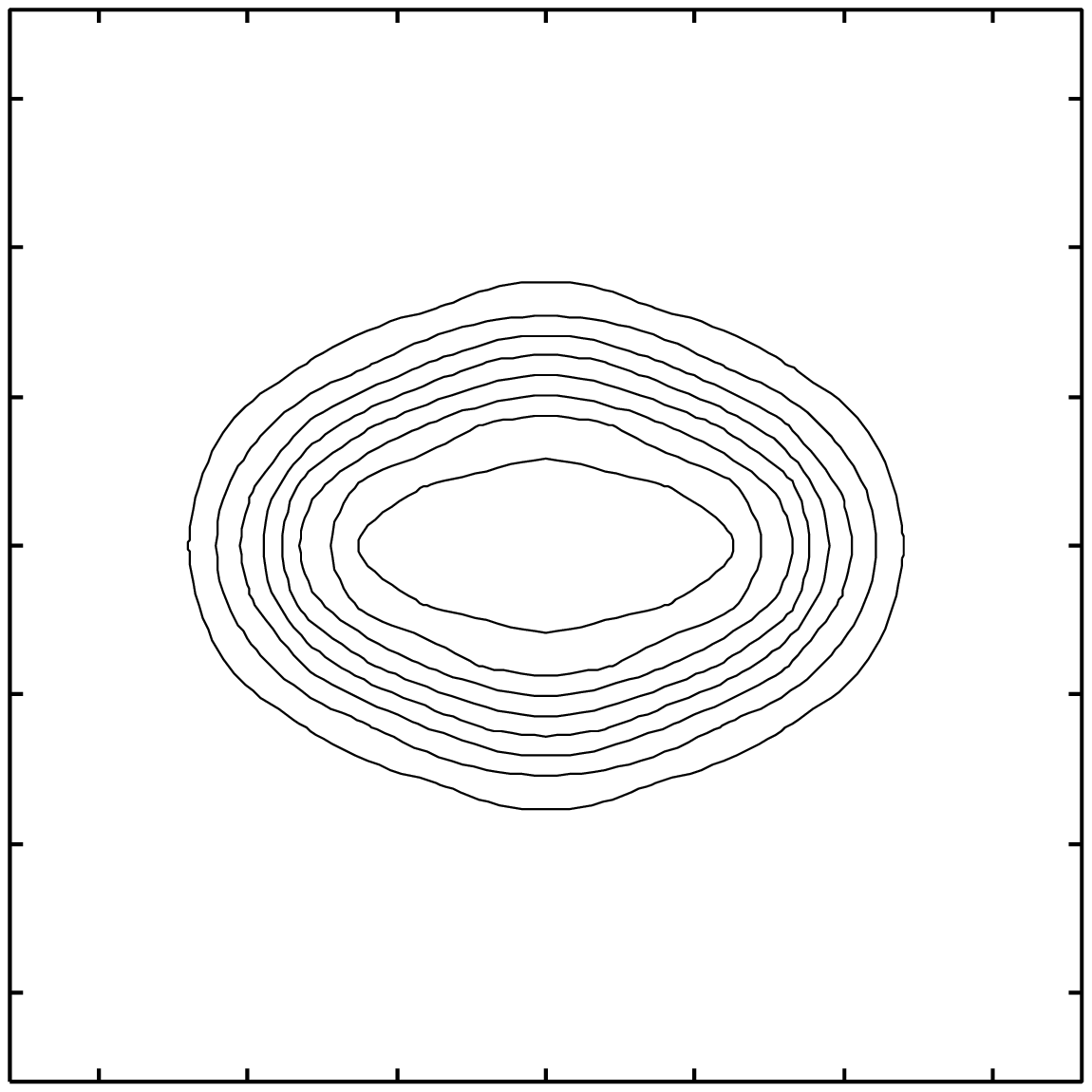}
\end{minipage}
\\
\begin{minipage}{.10\linewidth}
(b)
\end{minipage}
\begin{minipage}{.24\linewidth}
\vspace{0.2em}
\includegraphics[width=\linewidth]{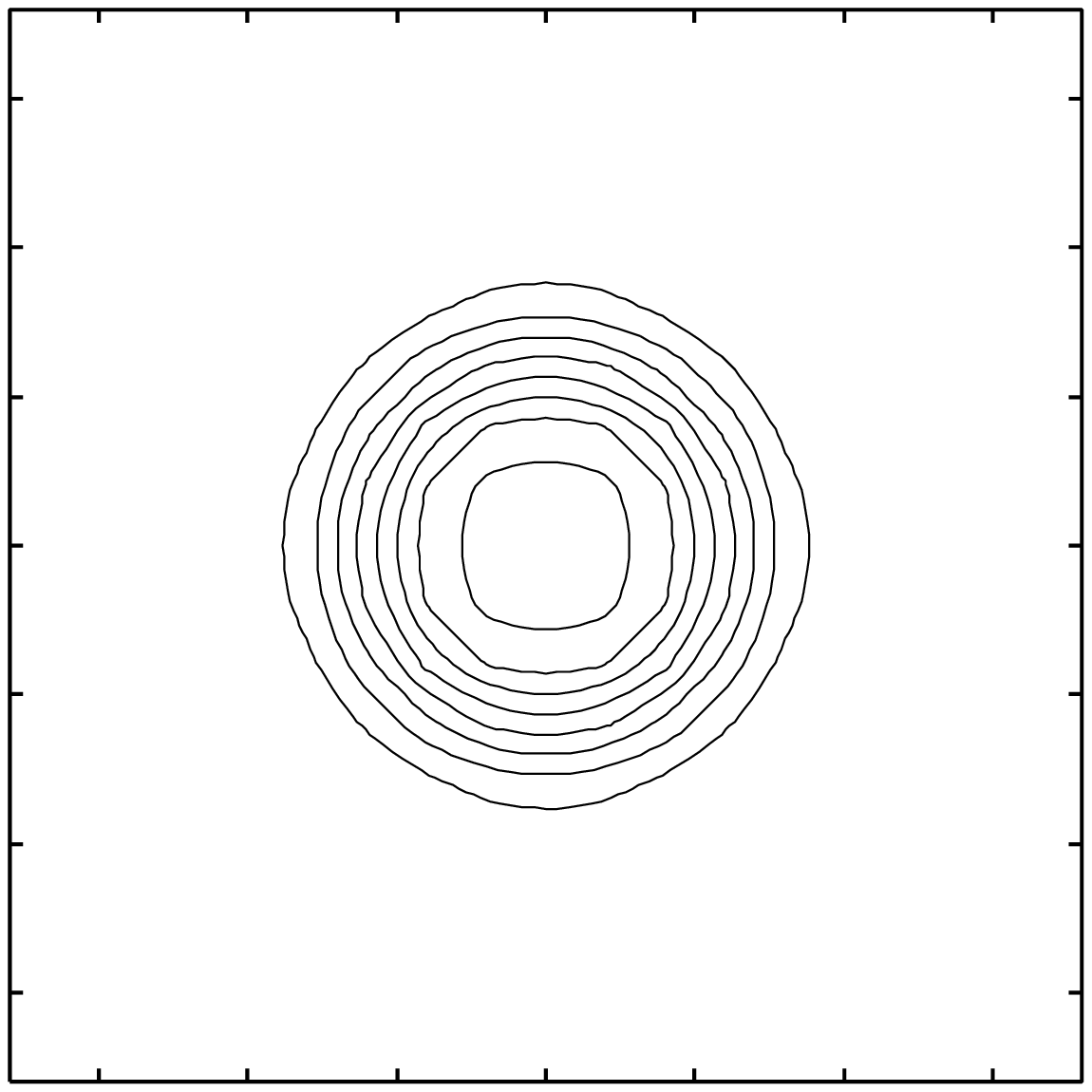}
\end{minipage}
\begin{minipage}{.24\linewidth}
\vspace{0.2em}
\includegraphics[width=\linewidth]{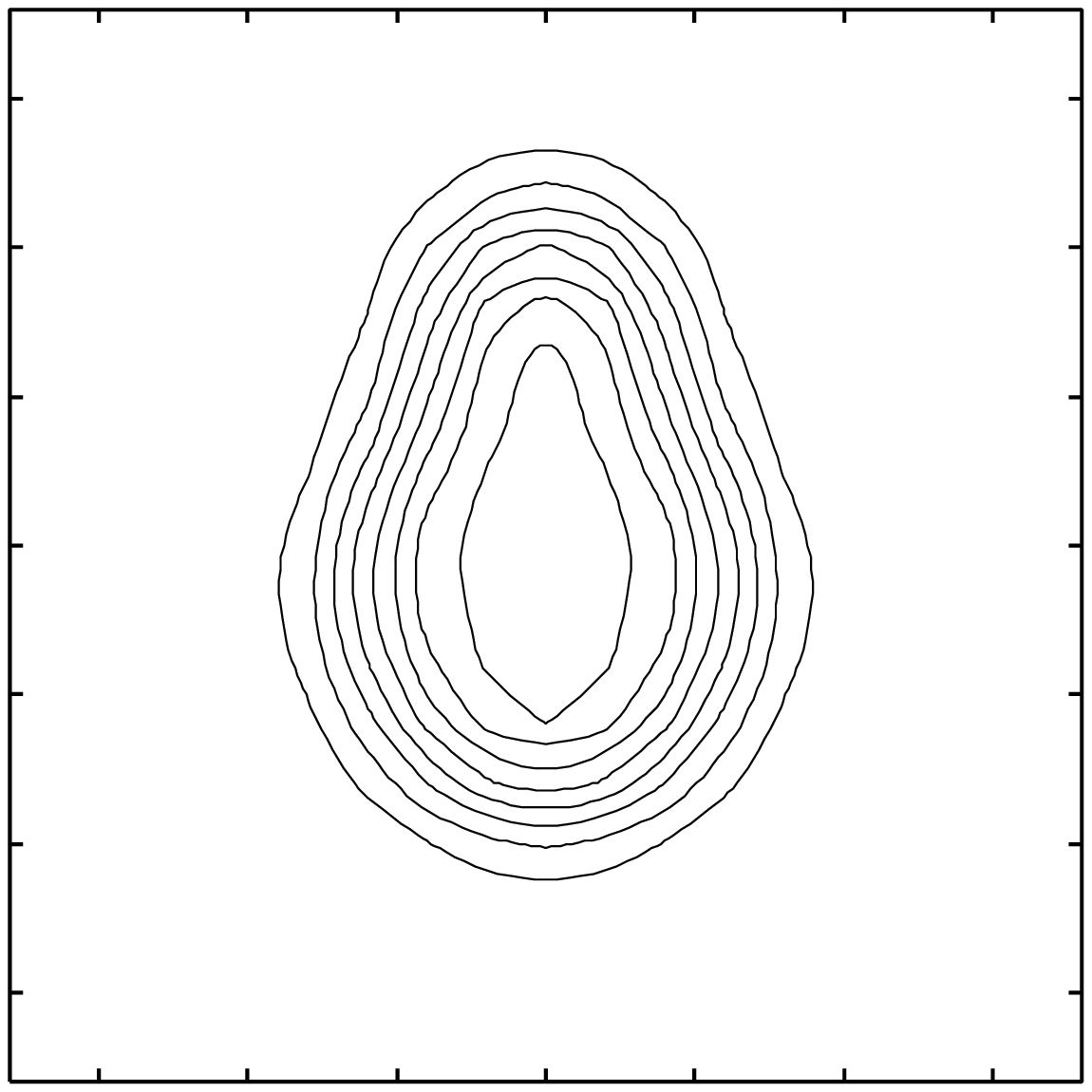}
\end{minipage}
\begin{minipage}{.24\linewidth}
\vspace{0.2em}
\includegraphics[width=\linewidth]{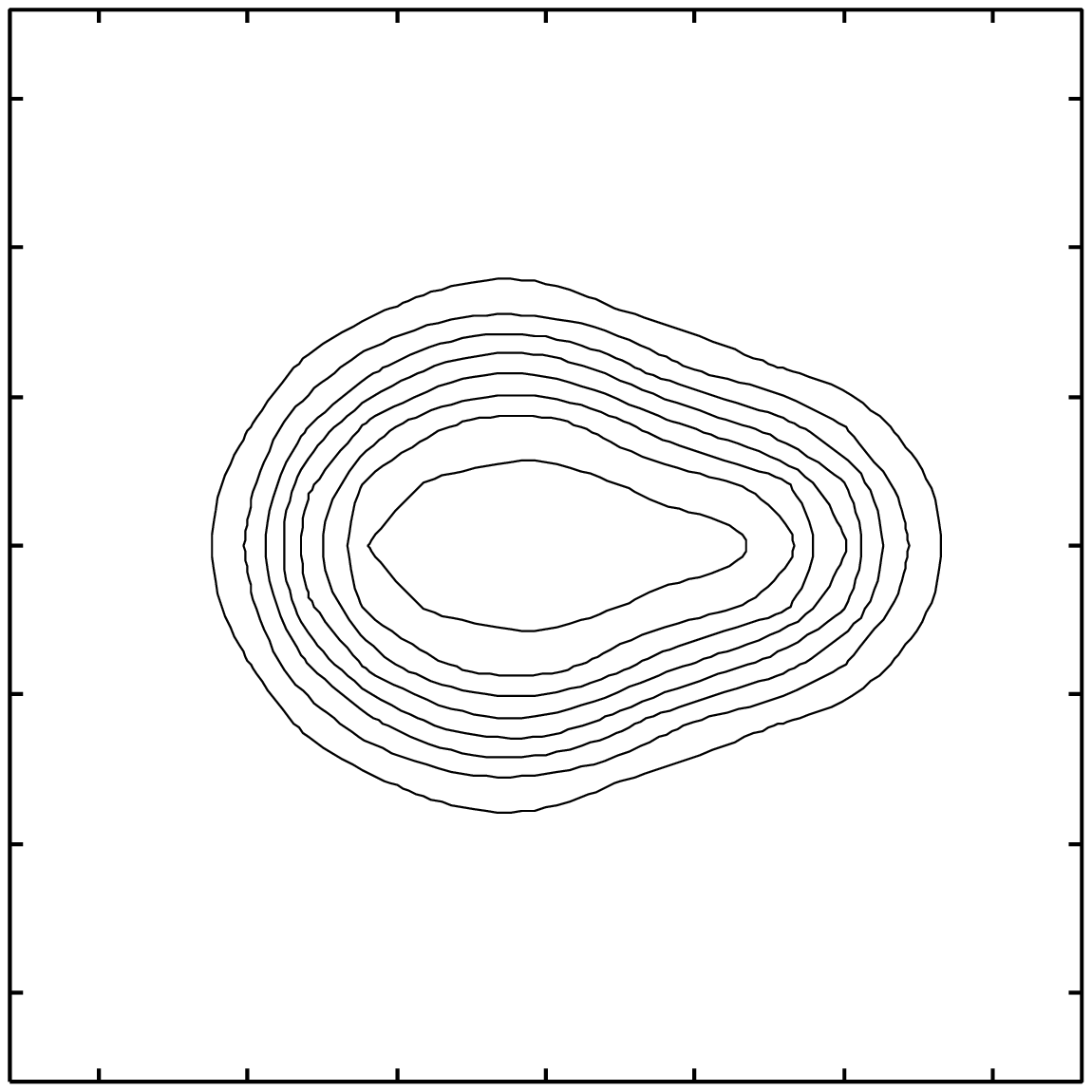}
\end{minipage}
\\
\begin{minipage}{.10\linewidth}
(c)
\end{minipage}
\begin{minipage}{.24\linewidth}
\vspace{0.2em}
\includegraphics[width=\linewidth]{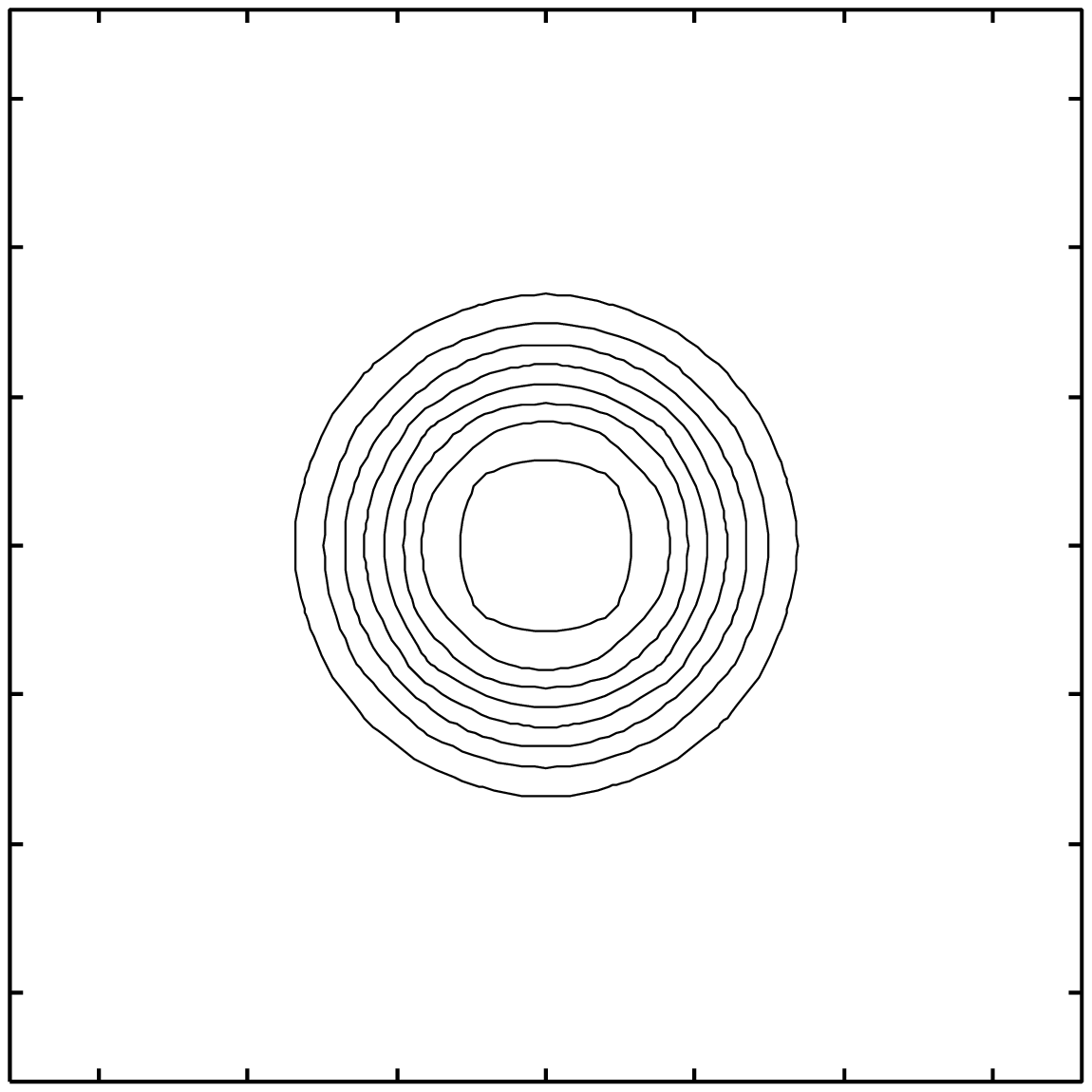}
\end{minipage}
\begin{minipage}{.24\linewidth}
\vspace{0.2em}
\includegraphics[width=\linewidth]{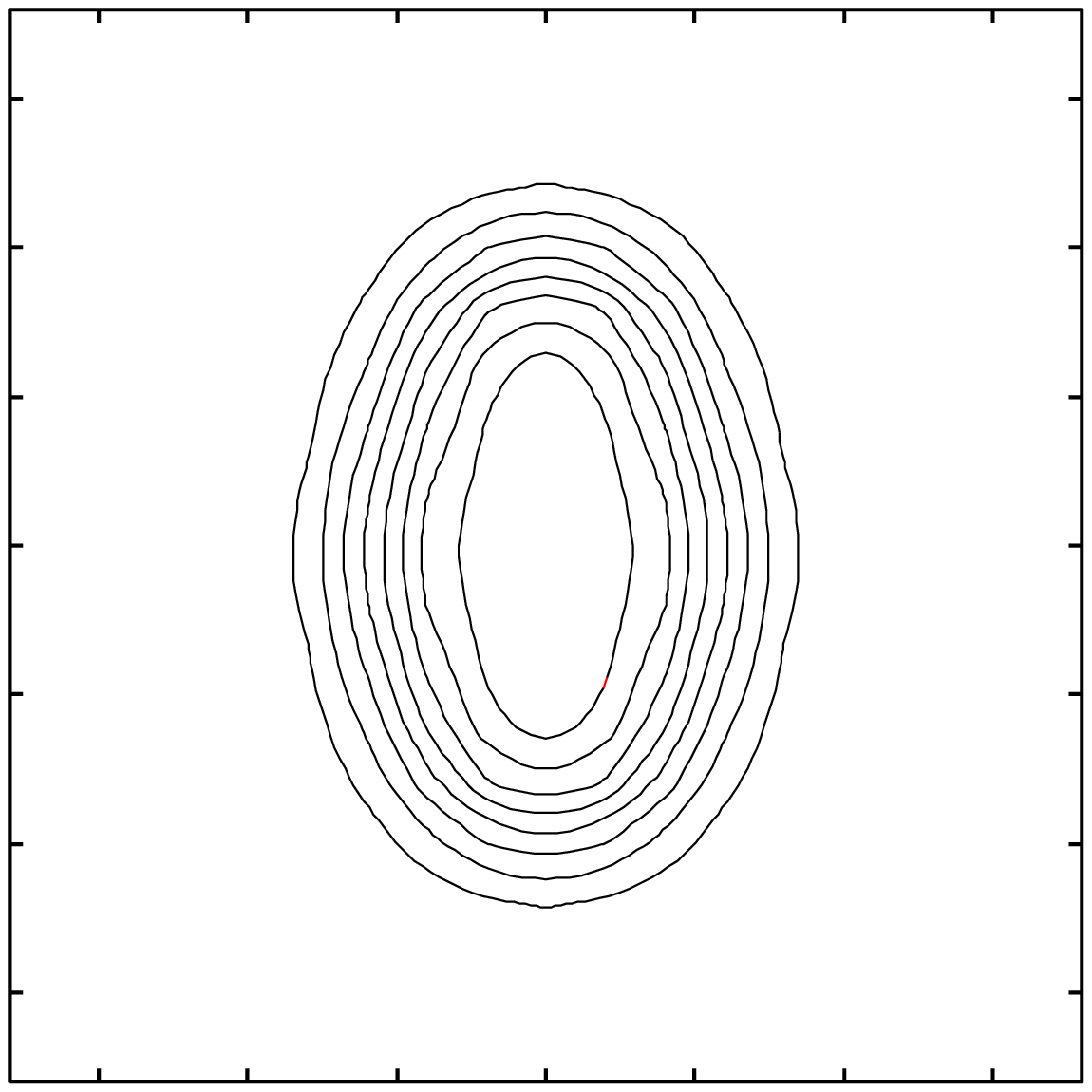}
\end{minipage}
\begin{minipage}{.24\linewidth}
\vspace{0.2em}
\includegraphics[width=\linewidth]{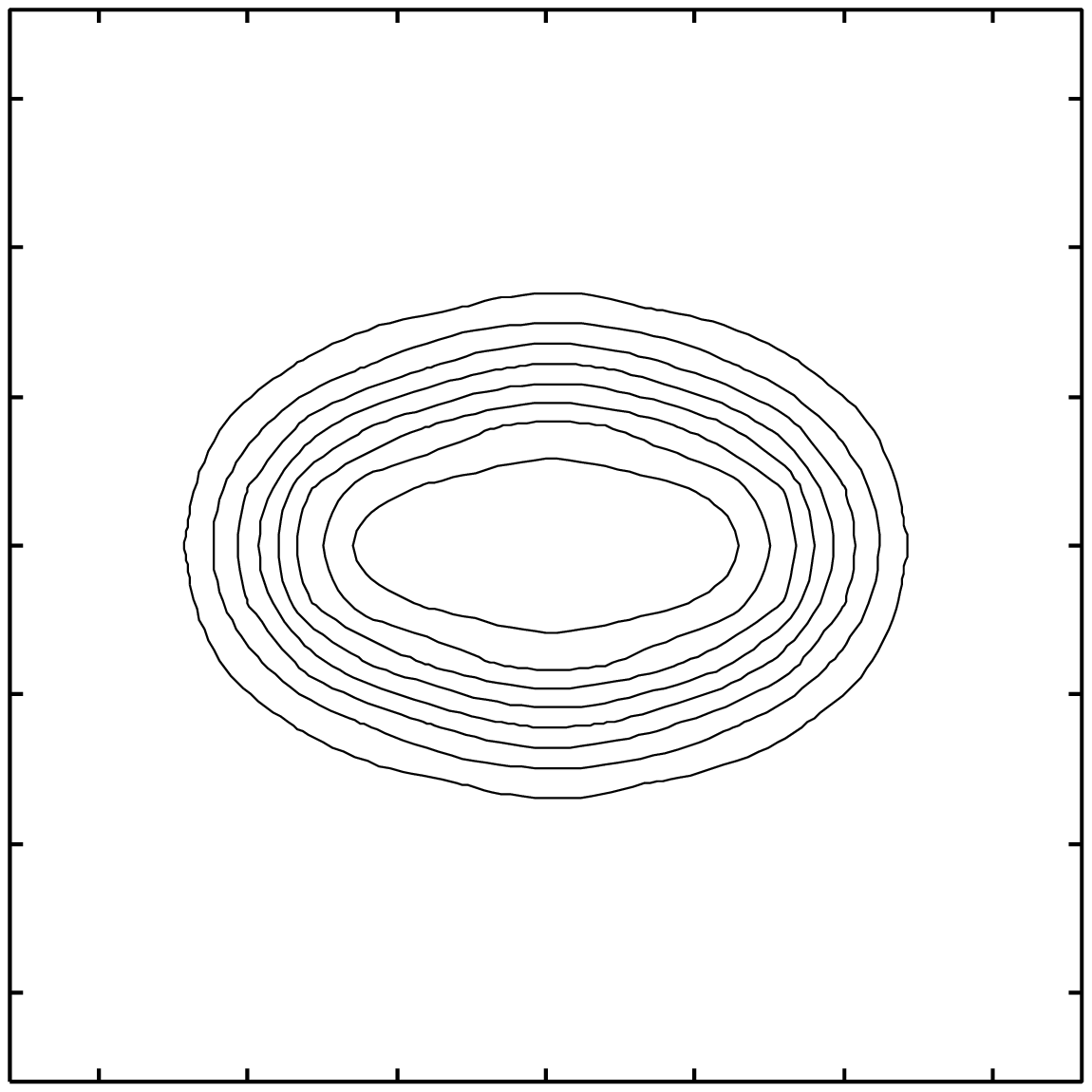}
\end{minipage}
\\
\begin{minipage}{.10\linewidth}
\hspace{0.5em}
(d)
\end{minipage}
\begin{minipage}{.2705\linewidth}
\vspace{0.23em}
\hspace{-1.05em}
\includegraphics[width=\linewidth]{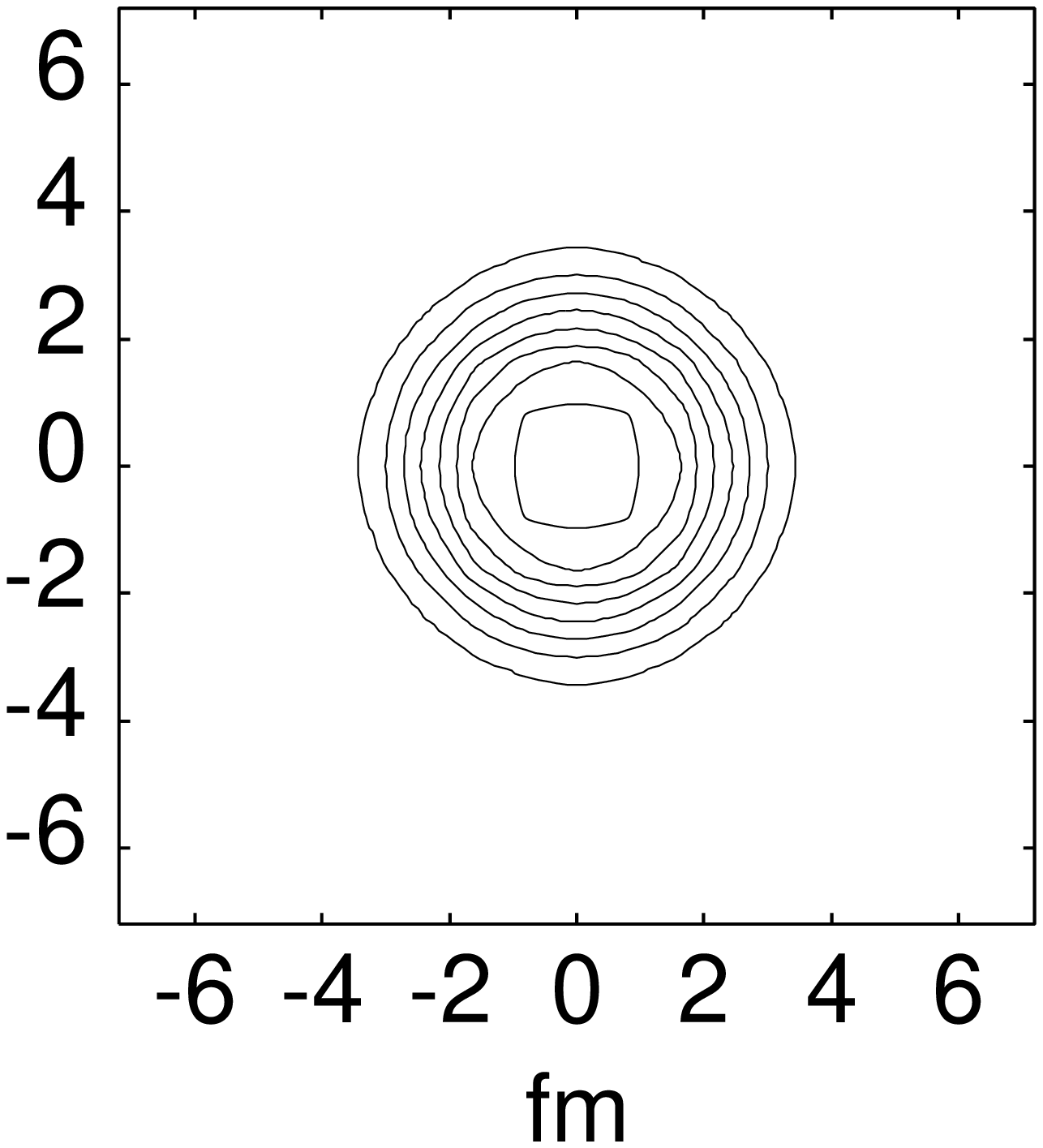}
\end{minipage}
\begin{minipage}{.24\linewidth}
\vspace{-1.407em}
\hspace{-1.15em}
\includegraphics[width=\linewidth]{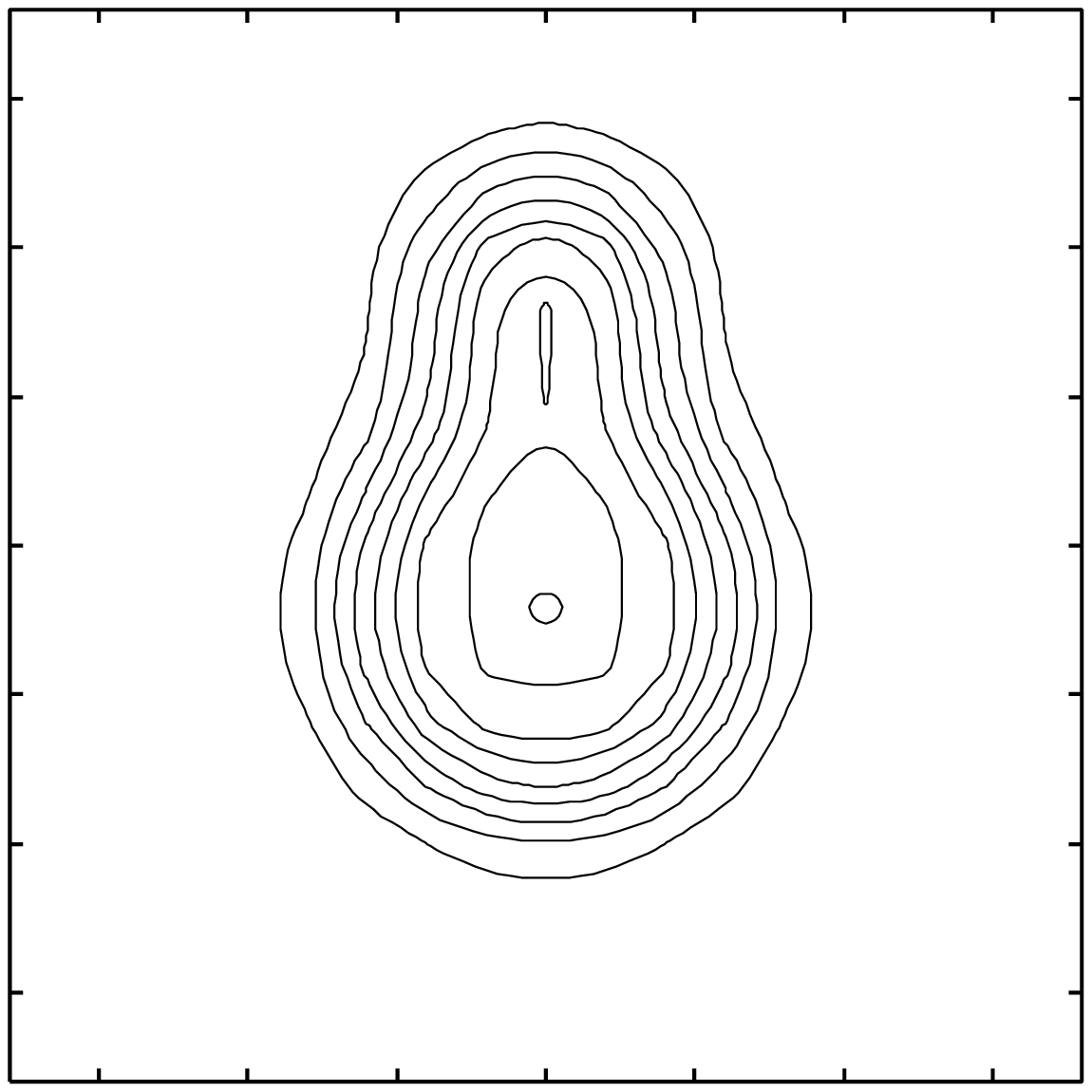}
\end{minipage}
\begin{minipage}{.24\linewidth}
\vspace{-1.407em}
\hspace{-1.15em}
\includegraphics[width=\linewidth]{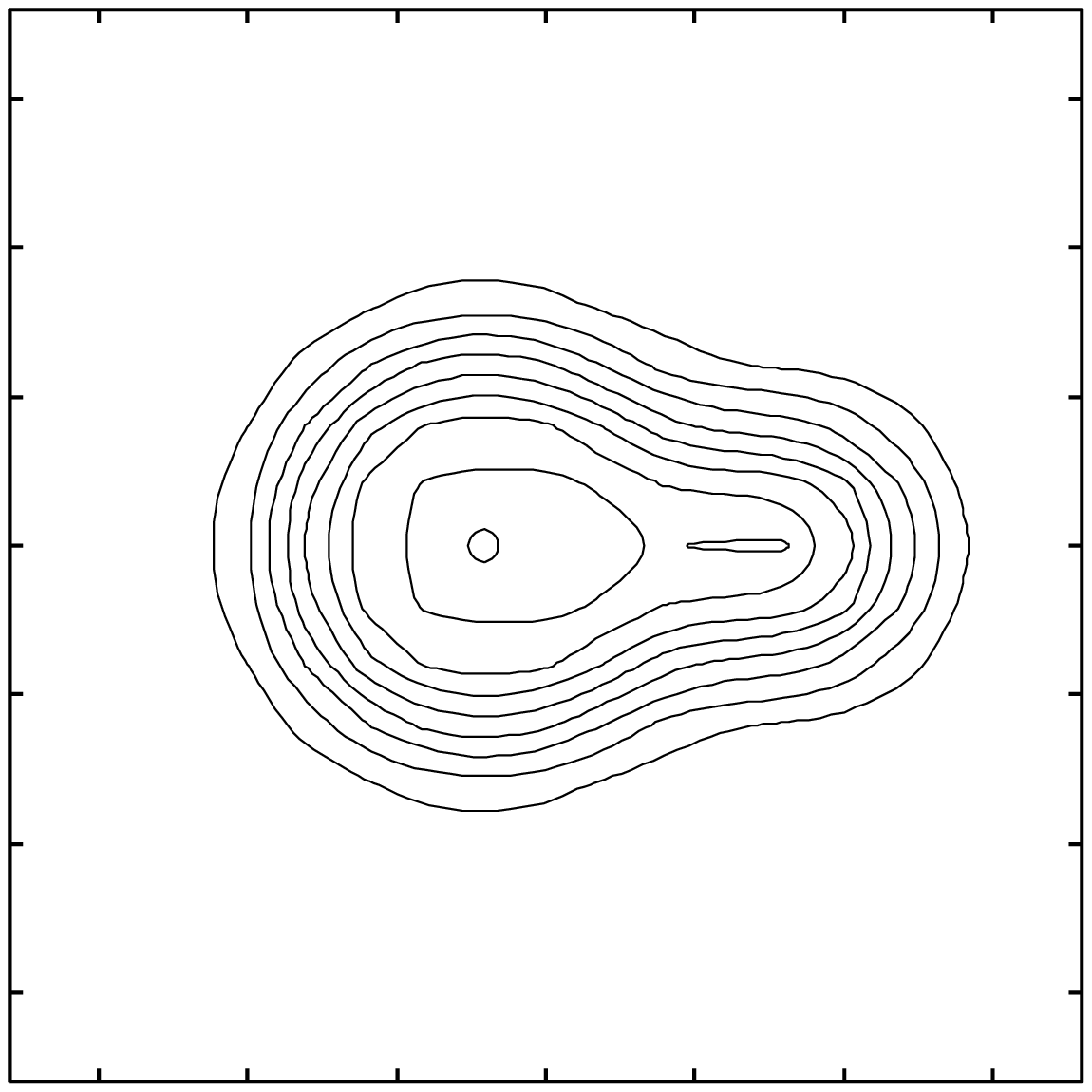}
\end{minipage}
\caption{\label{Ne20DENS}
Density distribution of the internal state in $^{20}$Ne,
in the $xy$, $yz$ and $zx$ planes where $x$, $y$ and $z$ 
axes represent the principal inertia axes, for
(a) ground $K^\pi=0^+$ state of the SHF calculation,
(b) ground $K^\pi=0^+$ state of PPSHF,
(c) excited $K^{\pi}=2^-$ state of PPSHF, and
(d) excited $K^{\pi}=0^-$ states of PPSHF, respectively.
The side of each panel is 14.4 fm. The contour lines are plotted for
 every 0.02 fm$^{-3}$. 
}
\end{figure}

Fig.~\ref{Ne20DENS} shows density distribution of the internal Slater
determinant $\Phi$ in the three planes which include two principal 
inertia axes.
The panels (a) are the SHF calculation, and the panels (b) are the
PPSHF for positive parity. They correspond to the ground state. 
The panels (c) and (d) are the PPSHF calculations for negative parity. 
The panel (c) is the lowest energy $K^{\pi}=2^-$ state, and (d) is 
the $K^{\pi}=0^-$. The nuclear shapes of (b) and (d) are 
characterized by the strong $\beta_{30}$ deformation and are 
considered to reflect the $\alpha$-$^{16}$O cluster structure. 
The clustering is stronger in the negative parity solution, 
which is consistent with the cluster model studies \cite{Fujiwara80}. 
The lowest energy negative-parity solution shown in (c) looks 
prolate although it has a small $\beta_{32}$ deformation.

As for the lowest negative parity state with $K^{\pi}=2^-$,
the excitation energy before AMP (the energy difference between 
the negative- and positive-parity solutions) is 5.47 MeV and that 
after AMP (the energy difference between $J^{\pi}=0^+$ and 
$J^{\pi}=2^-$ solutions in the parity and angular momentum 
projections) is 4.91 MeV. These values are close to the 
measured $2^-$ excitation energy of $^{20}$Ne, 4.97 MeV. 

The next negative parity solution with $K^{\pi}=0^-$, which shows 
large $\beta_{30}$ deformation, has $\alpha$-$^{16}$O cluster 
structure, as mentioned before.
The excitation energy before AMP is 7.96 MeV and that after AMP is 
6.42 MeV, which are slightly higher than the measured excitation
energy 5.78 MeV of the $J^{\pi}=1^-$ level.
The different character of the two negative-parity solutions with
$K^{\pi}=2^-$ and $K^{\pi}=0^-$ manifests clearly in the
spin-orbit energy, $E_{LS}$, the expectation value of the two-body 
spin-orbit force. The $K^{\pi}=2^-$ solution 
has large spin-orbit energy which is supposed to reflect 
dominant shell-model-like 5$p$-1$h$ configuration. 
The spin-orbit energy of $K^{\pi}=0^-$ solution is much smaller, 
even smaller than that in the ground state.
This is consistent with the development of $\alpha$-$^{16}$O 
cluster structure in the negative-parity states which was seen 
in the density distribution in Fig.~\ref{Ne20DENS}.

\begin{figure}[tb]
\includegraphics[scale=0.4]{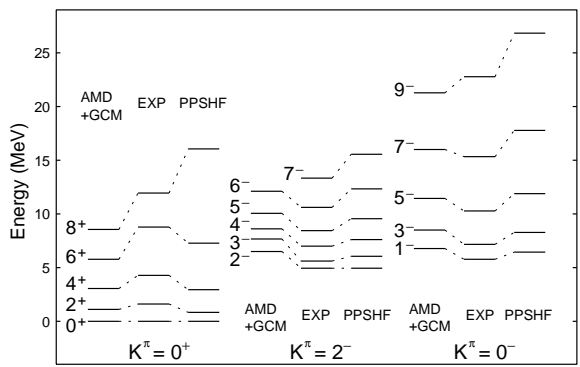}
\caption{\label{Ne20SPEC}
Energy spectra of $^{20}{\rm Ne}$ calculated with the AMP.}
\end{figure}

In Fig.~\ref{Ne20SPEC}, we show the excitation spectra of 
$^{20}$Ne after AMP. 
For three rotational bands of $K^{\pi}=0^+$, $2^-$, and $0^-$,
calculated spectra are compared with measurements and with the
results by the AMD (deformed AMD + GCM calculation with Gogny force) 
\cite{Kanada03}.
Since each band is well characterized by the $K$ quantum number, 
we simply show the energies given by 
$E^{J(\pm)}_K=h^{J(\pm)}_{KK}/n^{J(\pm)}_{KK}$.
Although the band head energies of $0^+$, $2^-$, and $1^-$ 
are described reasonably well, the calculated moment of inertia
deviates from the measured value. The calculated moment of inertia is
too large for the ground state band.
On the other hand, the calculated moment of inertia for the negative 
parity bands is slightly too small, opposite to the positive parity band.
The bandhead energy of the $K^\pi=2^-$ band in the AMD calculation is not
as good as the PPSHF, probably because the present work has a better
account of the single-particle wave functions.

The AMD method gives better description for the high spin levels, 
$6^+$ - $8^+$ energy difference. In our calculation, the AMP is
performed from a single intrinsic state, 
while a change of nuclear shape is shown as the 
angular momentum increases in the AMD calculation \cite{Kanada03}.
This indicates that, if we incorporate the cranking in the PPSHF framework, 
we might obtain better description for the higher angular momentum states.
The discrepancy in the moment of inertia looks similar between
PPSHF and AMD calculations. At present, we do not have a definite 
answer for the origin of this discrepancy.
The pairing correlation ignored in the present calculation may
be a possible answer.
Bender et al. calculated ground state bands of some light nuclei in the
HFBCS + GCM + AMP scheme \cite{Bender03-2}. Their calculation slightly
underestimates the moment of inertia, opposite to our result.
In the mean-field calculations, it has also been pointed out that the
moment of inertia is sensitive to the time-odd component which we
ignored \cite{Dobaczewski95}. 

In TABLE \ref{tab:20NeBE2}, we show observed and calculated intra-band
 B(E2) strengths of the $K^\pi=0^+$ and $K^\pi=2^-$ bands
 of $^{20}$Ne.
For the sake of comparison, we show results of the rotational model
 (Rot.) and the AMD + GCM \cite{Kanada03}.
Our calculation well reproduces B(E2) values of the 2$^-$ band.
For the 0$^+$ band, our calculation somewhat underestimates the B(E2) values
 of low angular momentum states.
It should be noted that we do not introduce any effective charge.
For $8^+ \rightarrow 6^+$ transition, our result overestimates the B(E2)
 value, and is close to the rotational model.
This is because we made AMP calculations from a single configuration.

\begin{table}[tb]
\caption{
\label{tab:20NeBE2}
Observed and calculated intra-band B(E2) strengths of the $K^\pi=0^+$
 and $K^\pi=2^-$ bands of $^{20}$Ne.
For the sake of comparison, we show results of the rotational model (Rot.)
 and the AMD + GCM \cite{Kanada03}.
The observed values are taken from Ref.~\cite{Hausser71}.
The values enclosed by curly brackets in the rotational model (rot.) are
 adjusted to experimental ones.
}
\begin{tabular}{l|cccc}\hline\hline
   $K^\pi=0^+$     & \hspace{0.5em} B(E2)$_{obs}$ & \hspace{1em} Rot. & AMD+GCM & PPSHF \\\hline
   $2^+ \rightarrow 0^+$ & \hspace{0.5em} 57 $\pm$ 8  & \hspace{1em} (57.0)     & 70.3       & 41.6        \\
   $4^+ \rightarrow 2^+$ & \hspace{0.5em} 71 $\pm$ 7  & \hspace{1em} 81.4       & 83.7       & 59.9        \\
   $6^+ \rightarrow 4^+$ & \hspace{0.5em} 66 $\pm$ 8 & \hspace{1em} 89.7       & 52.7       & 67.5        \\
   $8^+ \rightarrow 6^+$ & \hspace{0.5em} 24 $\pm$ 8  & \hspace{1em} 93.9       & 21.0       & 75.1        \\
  \hline\hline
   $K^\pi=2^-$     & \hspace{0.5em} B(E2)$_{obs}$ & \hspace{1em} Rot. & AMD+GCM & PPSHF \\\hline
   $3^- \rightarrow 2^-$ & \hspace{0.5em} 113 $\pm$ 29 & \hspace{1em}      101  & 102.8      & 97.6        \\
   $4^- \rightarrow 3^-$ & \hspace{0.5em}  77 $\pm$ 16 & \hspace{1em}       75  &  77.8      & 73.5        \\
   $4^- \rightarrow 2^-$ & \hspace{0.5em}  34 $\pm$ 6  & \hspace{1em}      (34) &  38.5      & 32.9        \\
   $5^- \rightarrow 4^-$ & \hspace{0.5em}  $<$ 808     & \hspace{1em}       53  &  84.5      & 52.6        \\
   $5^- \rightarrow 3^-$ & \hspace{0.5em}  84 $\pm$ 19 & \hspace{1em}       53  &  56.6      & 53.1        \\
   $6^- \rightarrow 5^-$ & \hspace{0.5em}  32 $\pm$ 13 & \hspace{1em}       39  &  29.9      & 39.7        \\
   $6^- \rightarrow 4^-$ & \hspace{0.5em}  55 $^{+23}_{-13}$ & \hspace{1em} 66  &  64.0      & 67.0        \\
  \hline\hline
\end{tabular}
\end{table}

\begin{figure}[tb]
\includegraphics[scale=0.4]{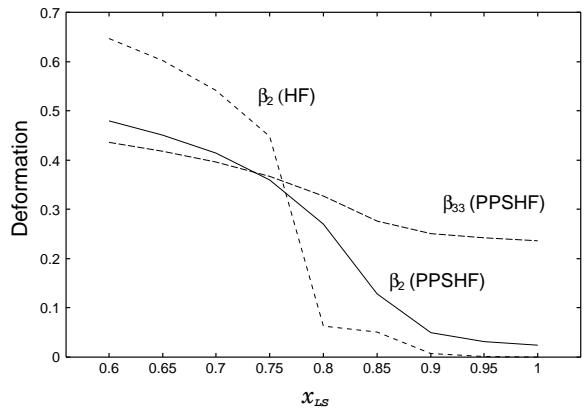}
\caption{\label{C12eDEFORM}
Deformation of $^{12}{\rm C}$ ground state
as a function of $x_{LS}$ which is a multiplicative factor
for the two-body spin-orbit interaction.
}
\end{figure}

We next discuss $^{12}$C.
In the SHF calculations, one usually obtain spherical ground
state for a $^{12}$C nucleus, although the rotational spectra is
observed experimentally. The spin-orbit interaction favors 
spherical structure to gain energy through $p_{3/2}$-$p_{1/2}$ 
splitting. If one weakens the spin-orbit interaction slightly, 
oblate deformation starts to appear in the ground state. 
We investigate change of the shape in the ground state by 
modifying the strength of the spin-orbit interaction,
multiplying a constant factor $x_{LS}$ to the two-body spin-orbit 
interaction of the Skyrme force. In Fig.~\ref{C12eDEFORM}, 
we show $\beta_2$ and $\beta_{33}$ of the positive parity solution 
as a function of $x_{LS}$. The $\beta_2$ value in the ordinary 
SHF calculation is shown as well.

\begin{table}[tb]
\caption{\label{tab:12C}
Calculated results of $^{12}$C at $x_{LS}=0.8$.
The meaning of the listed quantities are the same as those
in Table~\ref{tab:20Ne}.}
\begin{tabularx}{0.48\textwidth}{l|cXcc}
 \hline\hline
 & $K^\pi=0^+$ & \hspace{1em} negative parity & \hspace{-1em} SHF \\
 \hline
 Energy (EXP)    & $-92.162$     &\hspace{30pt}$+9.64$        & $-92.162$  \\
 \hline
 Energy (MF)     & $-94.617$     &\hspace{25pt}$+11.15$       & $-94.492$  \\
 Energy (MF+AMP) & $-96.499$     &\hspace{28pt}$+9.69$        &            \\
 ${\rm E_{LS}}$  & $-13.229$     &\hspace{25pt}$-7.550$       & $-15.191$  \\
 $\beta_2$       & $0.271$       &\hspace{30pt}$0.638$        & $0.069$    \\
 $\gamma$        & $59.95^\circ$ &\hspace{30pt}$27.77^\circ$  &$60.00^\circ$ \\
 $\beta_{31}$    & $0.0$         &\hspace{30pt}$0.056$        & $0.0$      \\
 $\beta_{33}$    & $0.327$       &\hspace{30pt}$0.518$        & $0.0$      \\
 \hline\hline
\end{tabularx}
\end{table}

We have found that, in the PPSHF calculation, an oblate shape
with substantial $\beta_{33}$ deformation (triangle shape) appears 
if one employs a slightly weak spin-orbit interaction. 
This may be regarded as the appearance of the 3$\alpha$ clustering structure. 
The $\beta_{33}$ deformation does not appear in the ordinary SHF calculation.
In addition, the oblate deformation starts to appear at larger
$x_{LS}$ value in the PPSHF calculation than in the SHF.
In the ordinary SHF calculation, the appearance of the
oblate deformation starts abruptly at about $x_{LS}=0.75-0.8$.
Note that, in order to
remedy the isotope-shift problem in Pb isotopes, the spin-orbit potential
should be weakened by about 30 percent from the ordinary one \cite{Reinhard95}.

\begin{figure}[tb]
\begin{minipage}{.10\linewidth}
(a)
\end{minipage}
\begin{minipage}{.24\linewidth}
\includegraphics[width=\linewidth]{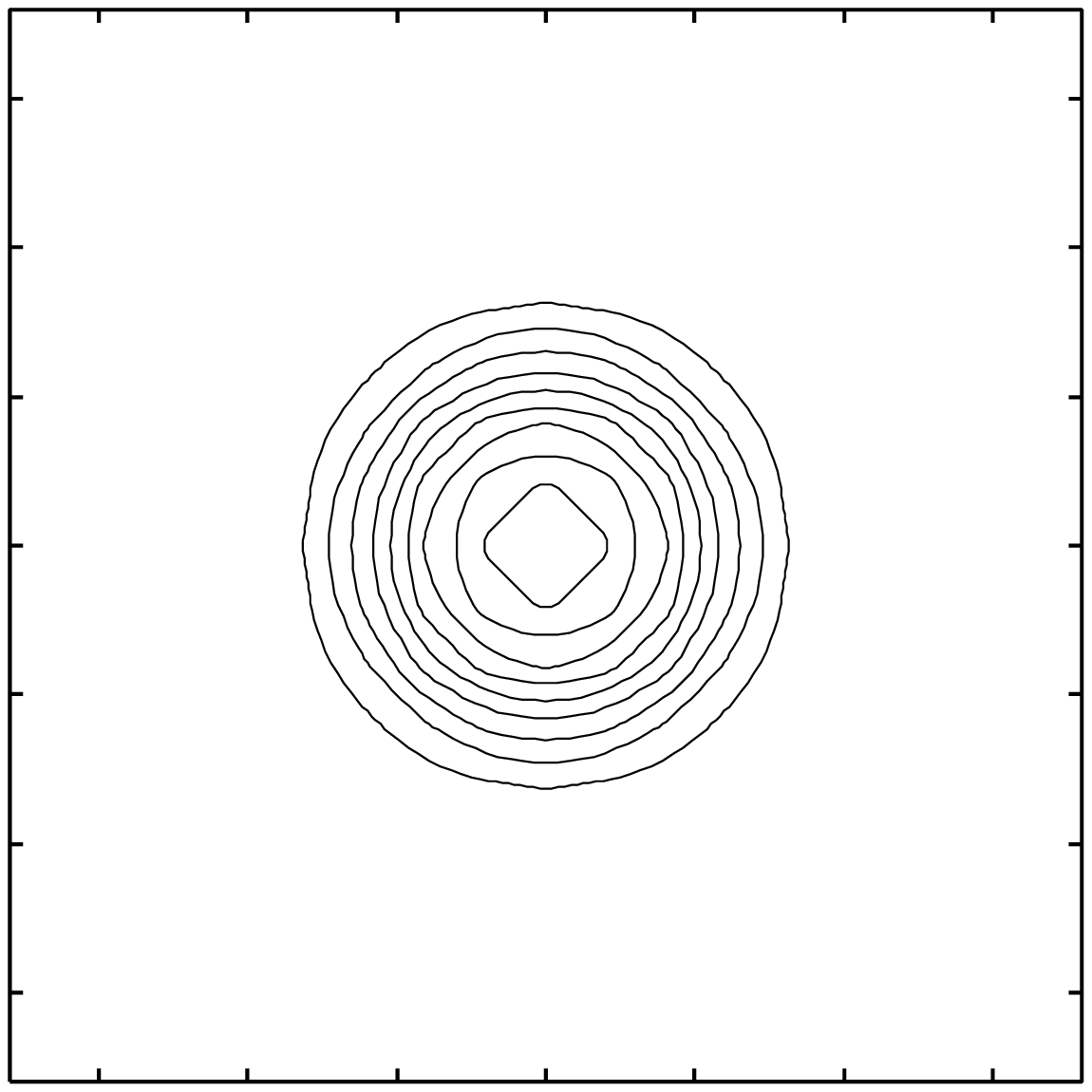}
\end{minipage}
\begin{minipage}{.24\linewidth}
\includegraphics[width=\linewidth]{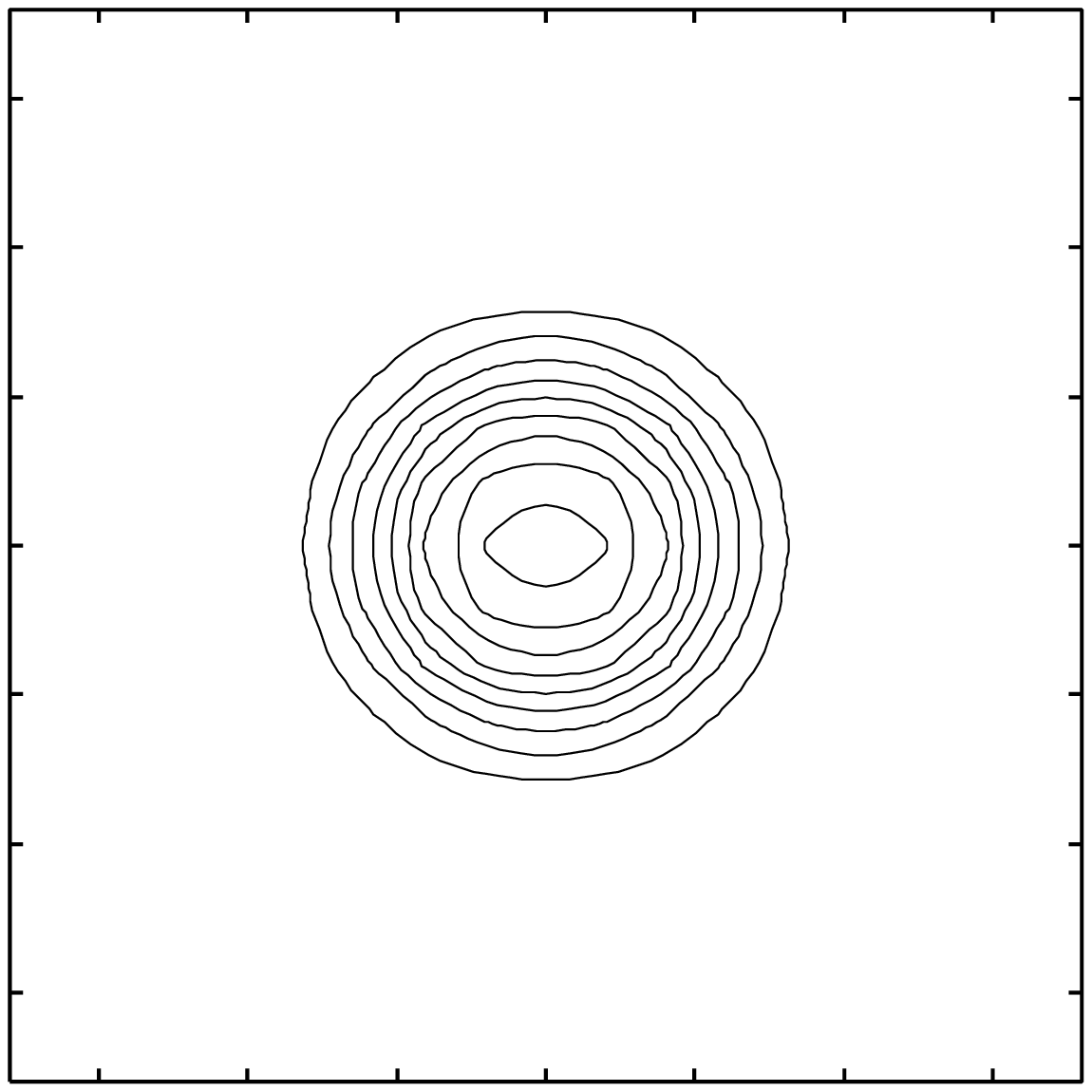}
\end{minipage}
\begin{minipage}{.24\linewidth}
\includegraphics[width=\linewidth]{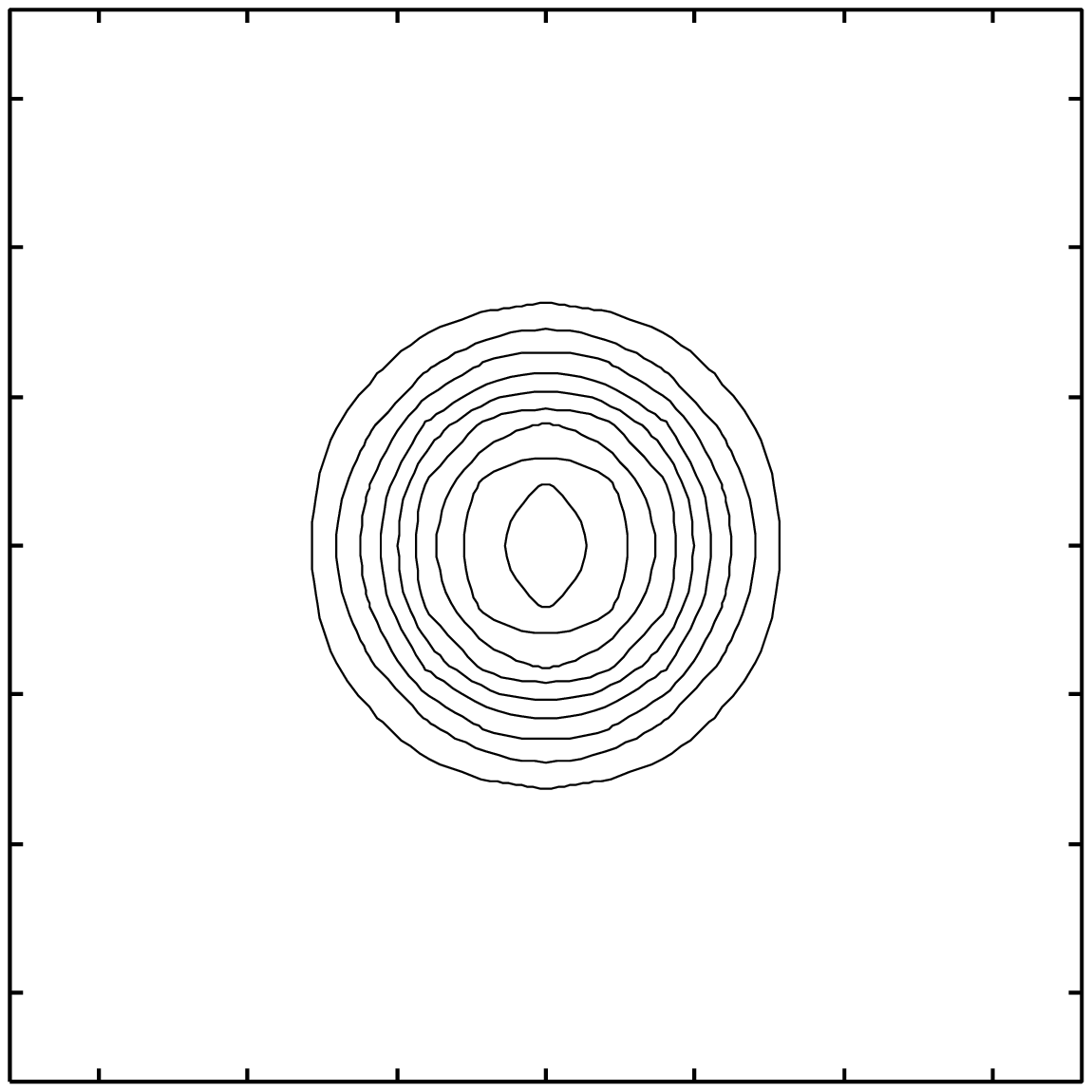}
\end{minipage}
\\
\begin{minipage}{.10\linewidth}
(b)
\end{minipage}
\begin{minipage}{.24\linewidth}
\vspace{0.2em}
\includegraphics[width=\linewidth]{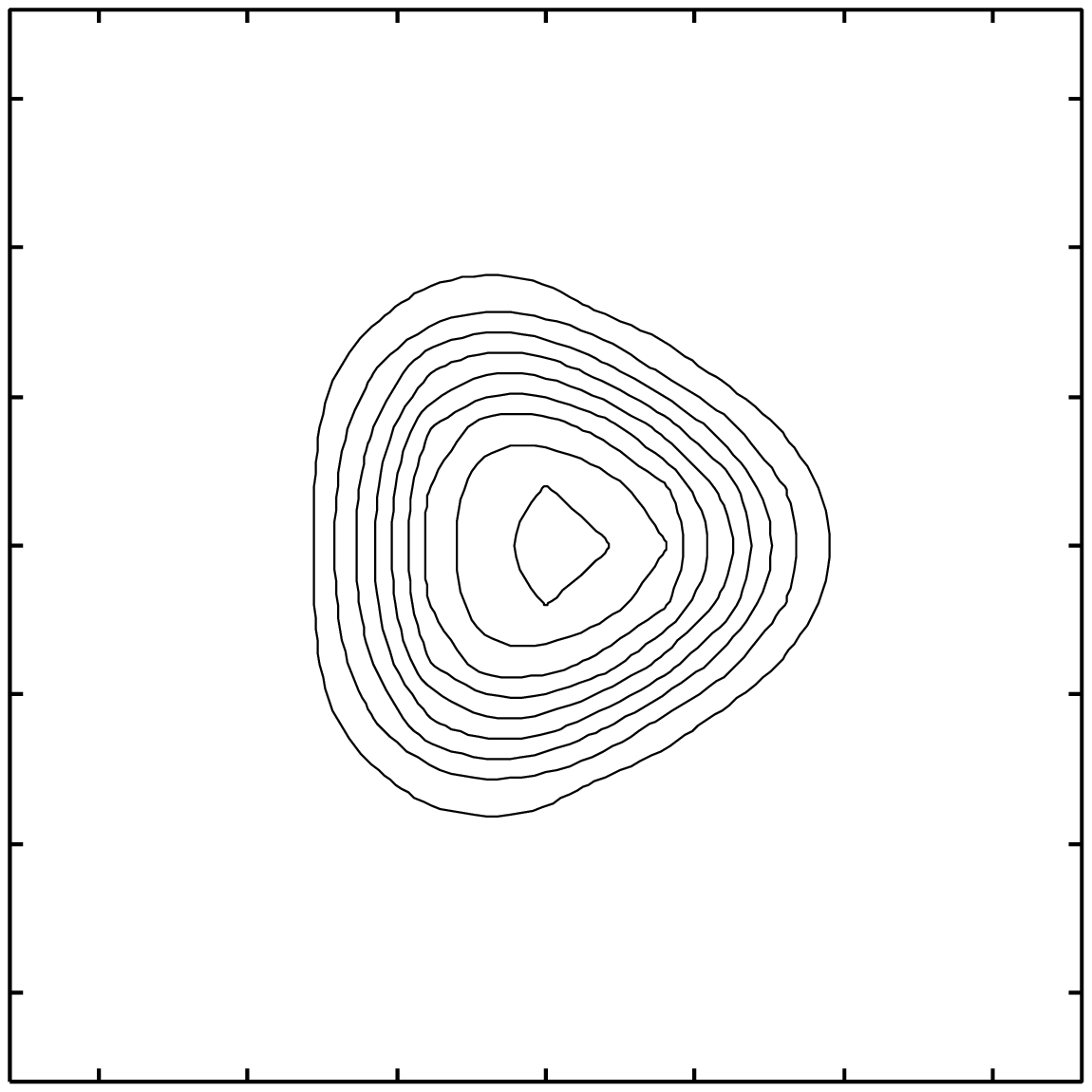}
\end{minipage}
\begin{minipage}{.24\linewidth}
\vspace{0.2em}
\includegraphics[width=\linewidth]{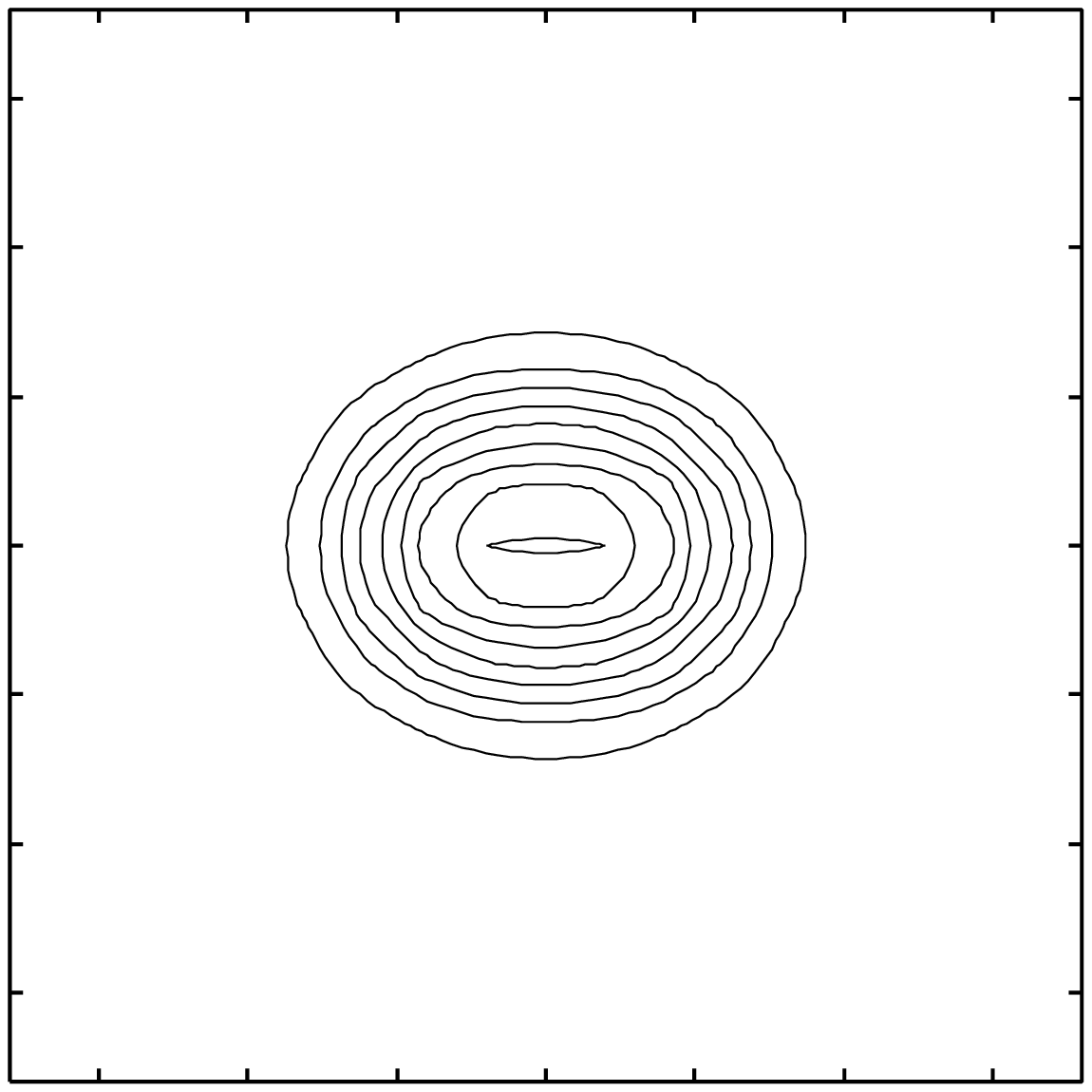}
\end{minipage}
\begin{minipage}{.24\linewidth}
\vspace{0.2em}
\includegraphics[width=\linewidth]{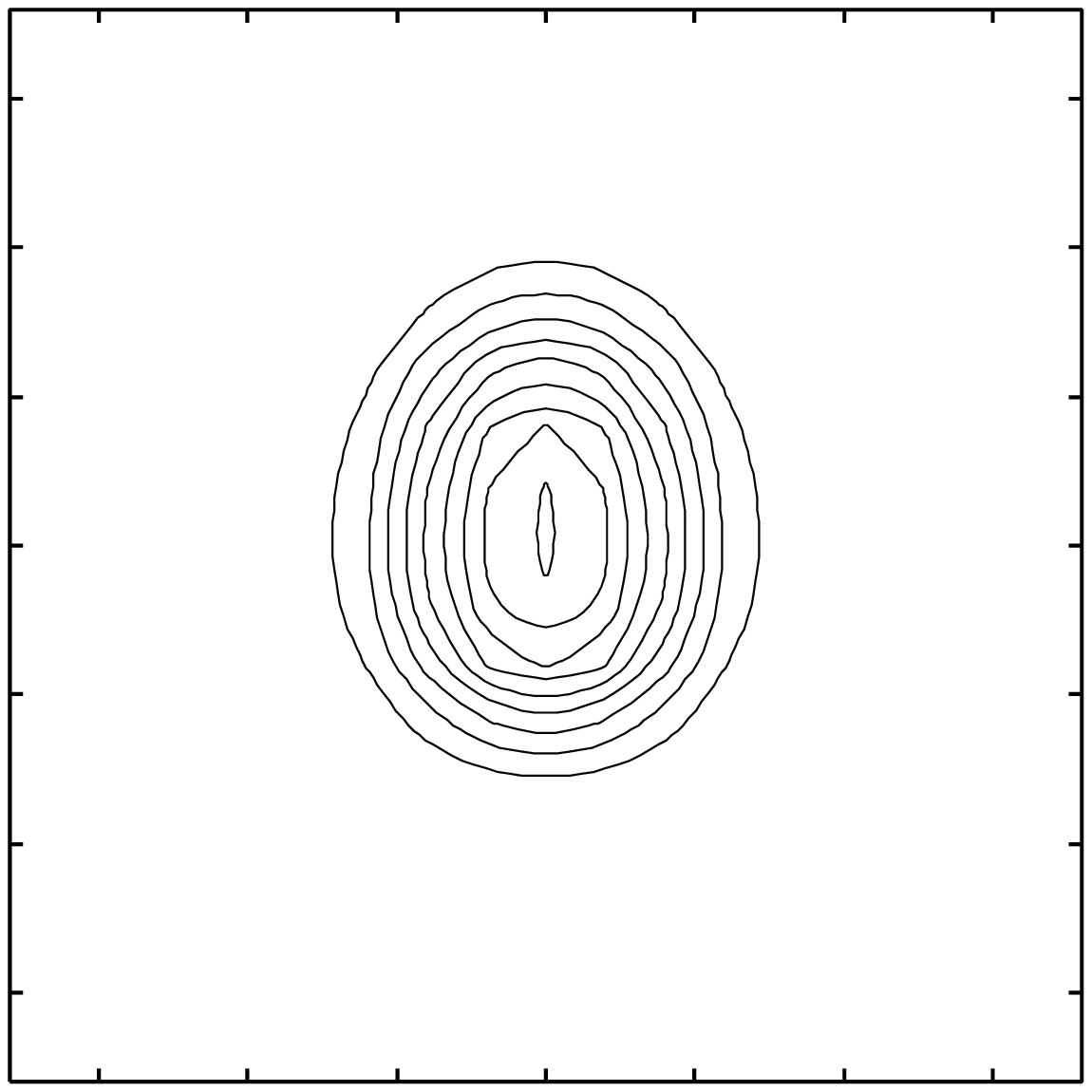}
\end{minipage}
\\
\begin{minipage}{.10\linewidth}
(c)
\end{minipage}
\begin{minipage}{.2705\linewidth}
\vspace{0.23em}
\hspace{-1.05em}
\includegraphics[width=\linewidth]{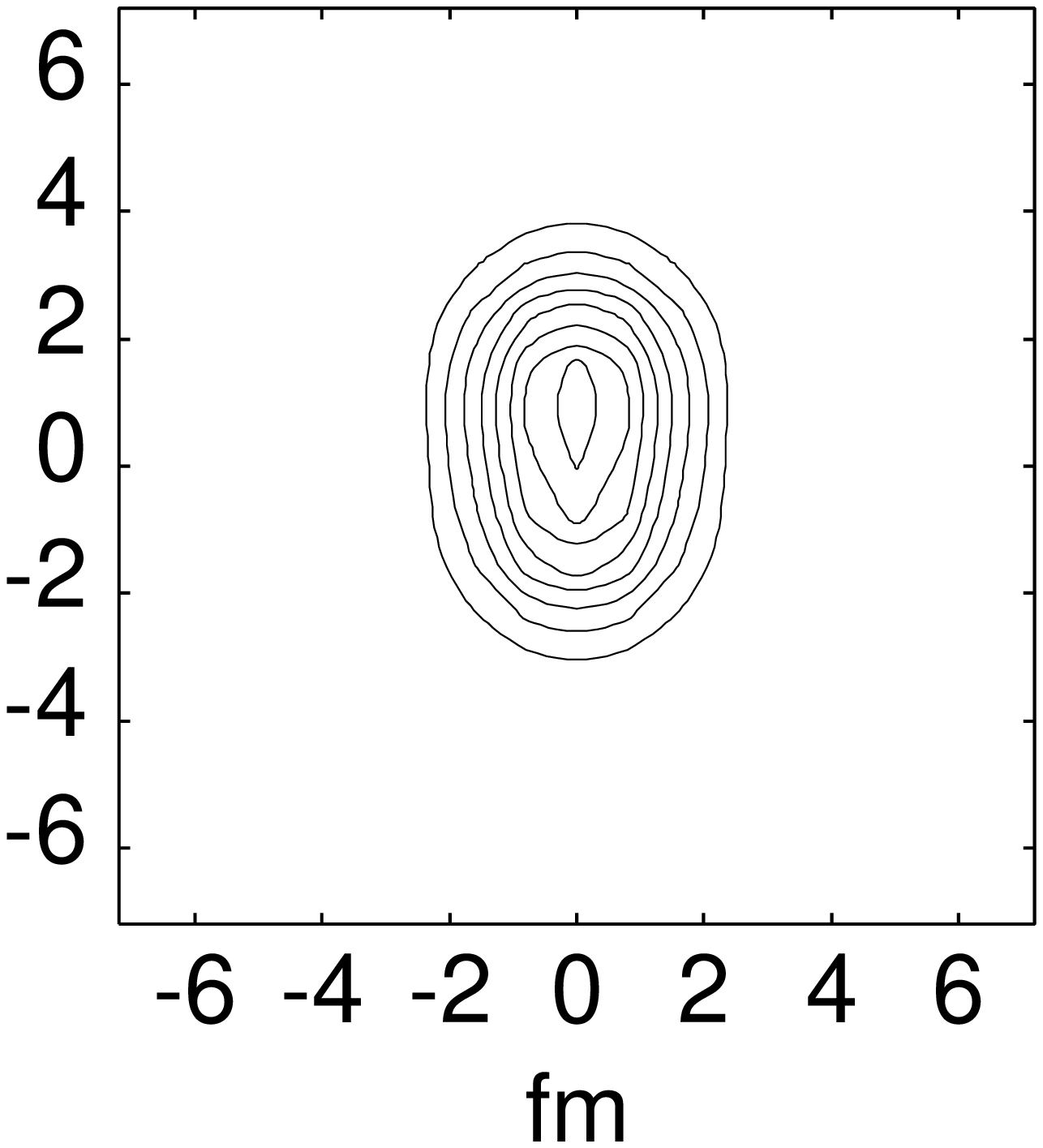}
\end{minipage}
\begin{minipage}{.24\linewidth}
\vspace{-1.407em}
\hspace{-1.15em}
\includegraphics[width=\linewidth]{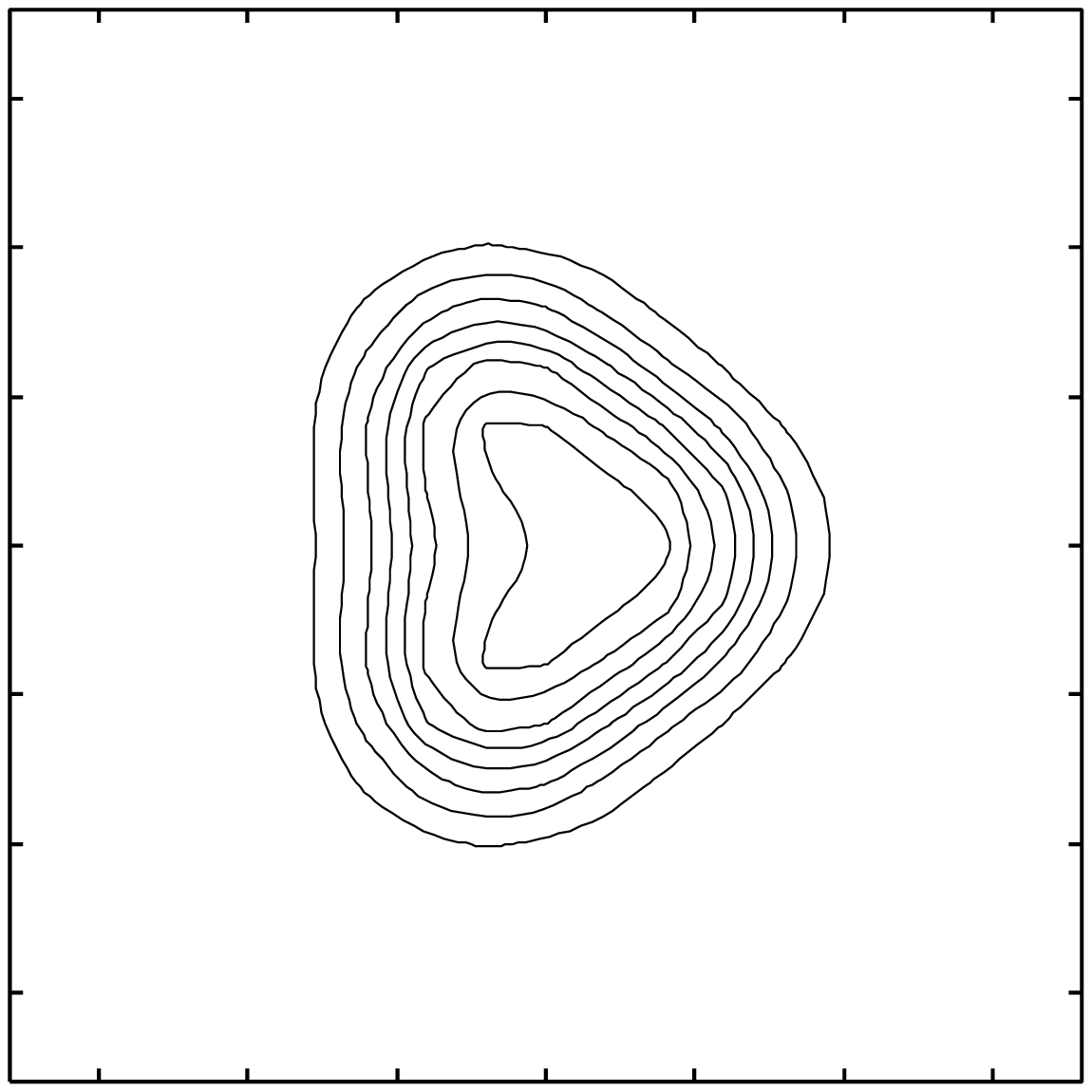}
\end{minipage}
\begin{minipage}{.24\linewidth}
\vspace{-1.407em}
\hspace{-1.15em}
\includegraphics[width=\linewidth]{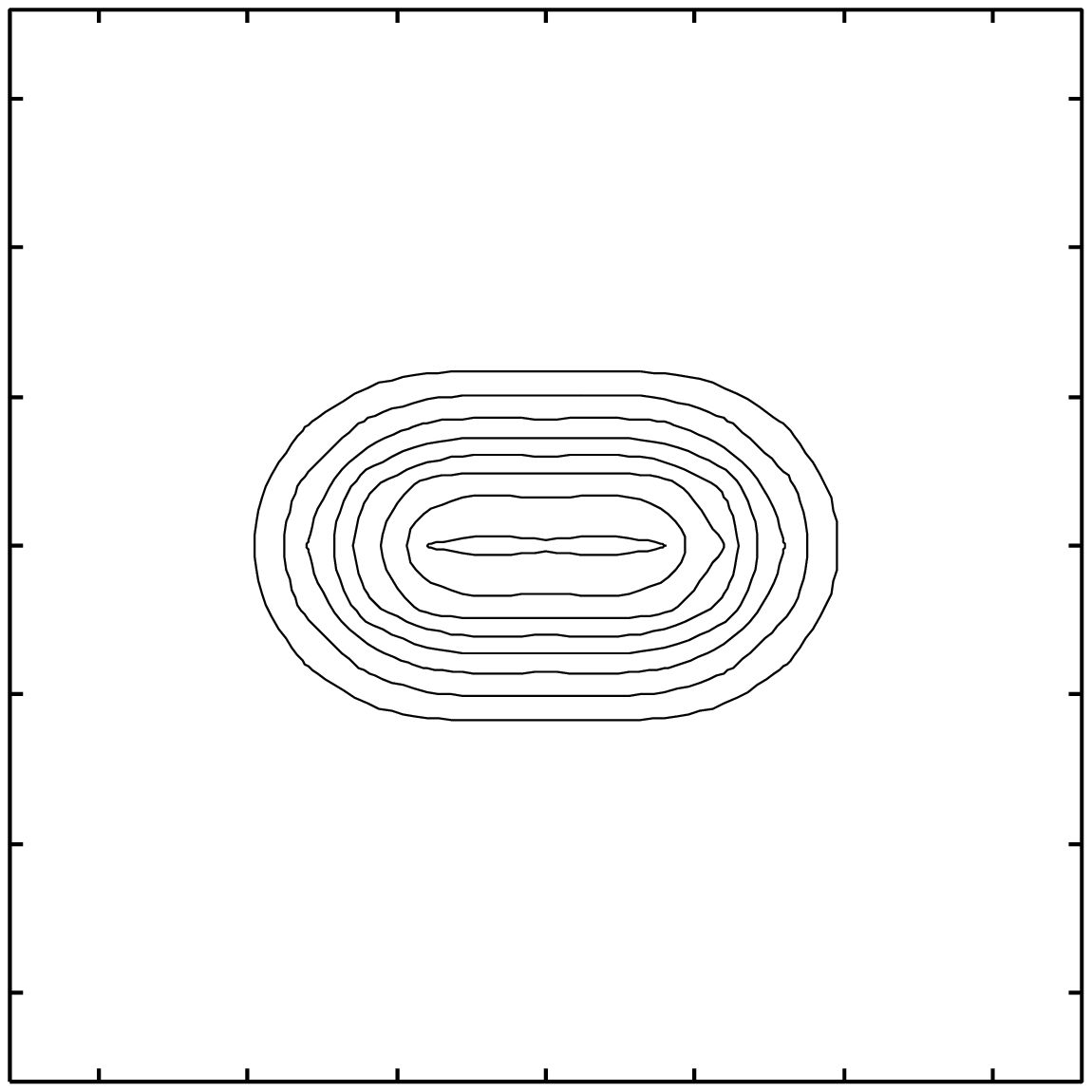}
\end{minipage}
\caption{\label{C12DENS}
Density distribution of the internal state in $^{12}{\rm C}$
at $x_{LS}=0.8$
for (a) SHF, (b) PPSHF ($K^\pi=0^+$), and
(c) PPSHF (negative-parity).
See the caption of Fig.~\ref{Ne20DENS} for explanation.}
\end{figure}

\begin{figure}[tb]
\includegraphics[scale=0.4]{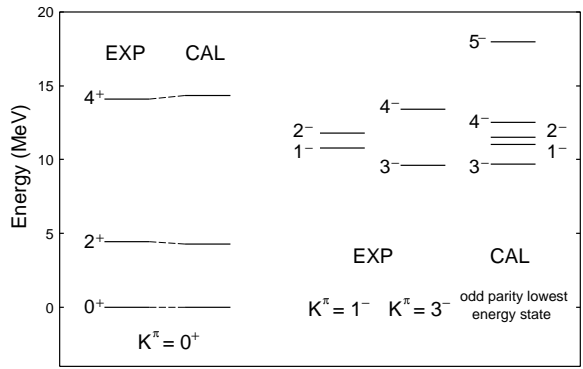}
\caption{\label{C12SPEC}
Energy spectra of $^{12}{\rm C}$ in the AMP.}
\end{figure}

We have examined AMP for solutions with different $x_{LS}$ values,
and have found that the ground state rotational band of $0^+,2^+,4^+$ 
is well described if we employ $x_{LS}$=0.8. At this value of $x_{LS}$,
we also made a calculation for negative parity solution.
We show below the results with this strength of the spin-orbit
interaction. The results are summarized for energies and deformations
in Table~\ref{tab:12C}, for intrinsic density distribution in 
Fig.~\ref{C12DENS},
and for the energy spectra in Fig.~\ref{C12SPEC}.
Calculated $B$($E2$:$~2^+ \!\! \rightarrow \! 0^+$) value is 5.14 e$^2$fm$^4$
which should be compared with the observed value 7.8 $\pm$0.4 e$^2$fm$^4$ 
\cite{Ajzenberg75}.

As seen in Fig.~\ref{C12DENS}~(a),
the nuclear shape is almost spherical in the ordinary SHF.
At a slightly small $x_{LS}$ value, oblate shape appears
in the ordinary SHF calculation (Fig.~\ref{C12eDEFORM}). 
In spite of the strong $\beta_2$ and $\beta_{33}$ deformation,
the spin-orbit energy is still 
large in the PPSHF calculation, almost comparable to that in the 
spherical SHF calculation. This indicates that the closed $p_{3/2}$ 
configuration has still significant effects on the ground state.

The negative parity solution shows strong $\beta_2$ deformation
with triaxiality. It has also strong octupole deformation, mainly
$\beta_{33}$ with small mixture of $\beta_{31}$. 
Reflecting this mixture in shape, two configurations with different
$K$ quantum numbers also coexist in the solution.
The panel (c)
of Fig.~\ref{C12DENS} shows that the 3$\alpha$ clustering is
much more developed in the negative parity than in the ground
state.
The AMP calculation gives reasonable description for excitation energy.
Because of $\gamma$ deformation, not only $3^-$ and $4^-$ states
 but also $1^-$ and $2^-$ states appear as the side band.

It turns out that the structure of the negative parity solution in the PPSHF 
depends strongly on the $x_{LS}$ value. 
If one employs the original Skyrme parameterization ($x_{LS}=1.0$), 
we obtain $K^{\pi}=1^-$ solution with $\beta_{31}$ deformation. 
On the other hand, for $x_{LS}$ values smaller than 0.7, we obtain 
$K^{\pi}=3^-$ solution with $\beta_{33}$ deformation. 
At $x_{LS}=0.8$, these two configurations mix up in the solution. 
Therefore, two distinct configurations coexist in the negative parity
at the excitation energies of about 10 MeV. In contrast to the $^{20}$Ne
case, two configurations do not separate but mix up.
In such a situation, treatments beyond the present framework seem
to be necessary, for example superposing multiple configurations in the 
generator coordinate treatment. We leave such an advanced treatment
as a future problem.

In summary, we propose the PPSHF method as a useful tool
to study both excited and ground states simultaneously.
Self-consistent solutions of the excited states can be obtained
in the negative parity, while the positive parity solutions 
describe the ground state incorporating certain correlation 
effects beyond the simple mean-field treatment.
We show the feasibility of such calculations employing the
uniform grid representation in the 3D Cartesian coordinate
and achieving the three-dimensional angular momentum projection.
The application to two $N=Z$ nuclei, $^{20}$Ne and $^{12}$C,
reveals that the obtained solutions show interesting deformations
violating reflection symmetries and incorporating clustering
correlation. We will apply, for the future, the present framework for the
systematic investigation and predictions of light nuclei, 
including exotic neutron rich nuclei.

This work is supported by the Grant-in-Aid for Scientific Research
(Nos. 14540369 and 14740146). A part of the numerical calculations
is achieved on the supercomputer at the Research Center for
Nuclear Study (RCNP), Osaka University.


\end{document}